\documentclass[aps,12pt,showpacs,preprint,groupedaddress,floatfix]{revtex4}
\usepackage{graphicx}
\usepackage{dcolumn}
\usepackage{bm}
\usepackage{amssymb}
\usepackage{amsmath}

\newcommand{\leftg}{\langle \phi_0 |}
\newcommand{\rightg}{| \phi_0 \rangle}

\begin{document}

\title{Relativistic Nuclear Energy Density Functionals: 
adjusting parameters to binding energies}
\author{T. Nik\v si\' c}
\author{D. Vretenar}
\affiliation{Physics Department, Faculty of Science, University of Zagreb, 
10000 Zagreb, Croatia}
\author{P. Ring}
\affiliation{Physik-Department der Technischen Universit\"at M\"unchen, 
D-85748 Garching,
Germany}
\date{\today}

\begin{abstract}
We study a particular class of relativistic nuclear energy density functionals in which 
only nucleon degrees of freedom are explicitly used in the construction of effective 
interaction terms. Short-distance (high-momentum) correlations, as well as 
intermediate and long-range dynamics, are encoded in the medium (nucleon 
density) dependence of the strength functionals  of an effective interaction Lagrangian. 
Guided by the density dependence of microscopic nucleon self-energies in nuclear 
matter, a phenomenological ansatz for the density-dependent coupling functionals is 
accurately determined in self-consistent mean-field calculations of binding energies 
of a large set of axially deformed nuclei. The relationship between the nuclear matter 
volume, surface and symmetry energies, and the corresponding predictions for  
nuclear masses is analyzed in detail. The resulting best-fit parametrization of the 
nuclear energy density functional is further tested in calculations of properties of 
spherical and deformed medium-heavy and heavy nuclei, including 
binding energies, charge radii, deformation parameters, neutron skin 
thickness, and excitation energies of giant multipole resonances.
\end{abstract}

\maketitle

\section{\label{secI}Introduction}
Among the microscopic approaches to the nuclear many-body
problem, the framework of nuclear energy density functionals (NEDF) provides 
the most complete and accurate description of ground-state properties and 
collective excitations over the whole nuclide chart. Probably
no other method achieves comparable accuracy at the same computational cost. 
At the level of practical applications the NEDF framework is realized in terms of 
self-consistent mean-field (SCMF) models based, for instance, on the Gogny 
effective interaction, the Skyrme energy functional, and the relativistic 
meson-exchange effective Lagrangian \cite{BHR.03,VALR.05}.
In the mean-field approximation the dynamics of the nuclear many-body system is 
represented by independent nucleons moving in self-consistent potentials, that  correspond to the actual density and current distributions of a given nucleus.

The SCMF approach to nuclear structure is analogous to Kohn-Sham density 
functional theory \cite{KS.65,Kohn.99}, and nuclear mean-field 
models approximate the exact energy functional, which includes all higher-order 
correlations, with powers and gradients of ground-state nucleon densities and 
currents \cite{LNP.641}. In particular, a number of very successful relativistic 
mean-field (RMF) models have been constructed based on the framework 
of quantum hadrodynamics (QHD) \cite{SW.86,SW.97}.
There are important advantages in using functionals with manifest 
covariance \cite{FS.00}. The most obvious is the natural inclusion of the 
nucleon spin degree of freedom, and the resulting nuclear spin-orbit potential 
which emerges automatically with the empirical strength in a covariant formulation. 
The consistent treatment of large, isoscalar, Lorentz scalar and vector self-energies 
provides a unique parametrization of time-odd components of the nuclear 
mean-field, i.e. nucleon currents, which is absent in the non-relativistic 
representation of the energy density functional. The empirical pseudospin 
symmetry in nuclear spectroscopy finds a natural explanation in terms of 
relativistic mean fields \cite{Joe.05}. On a microscopic level, it has been 
argued \cite{FS.00} that a covariant formulation of nuclear dynamics 
manifests the true energy scales of QCD in nuclei, and is consistent with 
the nonlinear realization of chiral symmetry through the implicit inclusion of 
pion-nucleon dynamics in the effective nucleon self-energies. A covariant 
treatment of nuclear matter provides a distinction between scalar and 
four-vector nucleon self energies, leading to a very natural saturation 
mechanism.

In conventional QHD the nucleus is described as a system of Dirac nucleons 
coupled to exchange mesons through an effective Lagrangian. 
The isoscalar scalar $\sigma$ meson,
the isoscalar vector $\omega$ meson, and the isovector vector $\rho$ meson
build the minimal set of meson fields that is necessary for a description of bulk 
and single-particle nuclear properties. Of course, at the scale of low-energy 
nuclear structure, heavy-meson exchange is just a convenient representation of 
the effective nuclear interaction. At the energy and momentum 
scales characteristic of nuclei, the only degrees of freedom that have to be 
taken into account explicitly in the description of many-body dynamics are 
pions and nucleons. The behavior of the nucleon-nucleon (NN) interaction at 
long and intermediate distances is determined by one- and two-pion exchange 
processes. The exchange of heavy mesons is associated with short-distance 
dynamics that cannot be resolved at low energies that 
characterize nuclear binding and, therefore, can be represented by 
local four-point (contact) NN interactions, with low-energy 
(medium-dependent) parameters adjusted to nuclear data. These concepts 
of effective field theory and density functional theory methods have recently 
been used to derive a microscopic relativistic energy density functional 
framework constrained by in-medium QCD sum rules and chiral 
symmetry \cite{FKV.04,FKV.06}. The
density dependence of the effective nucleon-nucleon couplings is determined
from the long- and intermediate-range interactions generated by one- and
two-pion exchange processes. They are computed using in-medium chiral
perturbation theory, explicitly including $\Delta(1232)$ degrees of freedom
\cite{FKW.05}. Regularization dependent contributions to the energy density of
nuclear matter, calculated at three-loop level, are absorbed in contact
interactions with parameters representing unresolved short-distance dynamics.

However, even in a fully microscopic approach that starts from a description 
of symmetric and asymmetric, homogeneous and inhomogeneous nuclear 
matter, the parameters of a nuclear energy density functional have still to be 
fine tuned to structure data of finite nuclei. This is simply because gross 
properties of infinite nuclear matter cannot determine the density functional 
on the level of accuracy that is needed for a quantitative description of 
structure phenomena in finite nuclei. For most functionals this tuning 
is performed on a relatively small set of spherical closed-shell nuclei, mainly 
because they are simple to calculate and can therefore be easily 
included in multiparameter {\em least-squares} fits. A problem arises, however, 
because ground-state data of closed-shell nuclei include long-range 
correlations that cannot really be absorbed into mean-field functionals. 
Generally this will affect the predictive power of energy density functionals 
when they are used in the description of phenomena related to the 
evolution of shell structure. 
For instance, soft potential energy surfaces and/or small energy differences 
between coexisting minima in deformed nuclei, are often difficult to describe 
using functionals adjusted solely to data of spherical nuclei, even when 
sophisticated models are employed that include angular momentum and 
particle number projection, as well as intrinsic configuration mixing. 

In this work we explore a class of relativistic energy density functionals 
originally introduced in Refs.~\cite{FKV.04,FKV.06} but, instead of using 
low-energy QCD constraints for the medium dependence of the parameters, 
a phenomenological ansatz is adjusted exclusively to masses of a relatively 
large set of axially deformed nuclei. The phenomenological approach, 
although guided by microscopic nucleon self-energies in nuclear matter, 
gives us more freedom to investigate in detail the relationship between global 
properties of a nuclear matter equation of state (volume, surface, and 
asymmetry energies) and the corresponding 
predictions for nuclear binding energies. Eventually the goal will be 
to develop an energy density functional that does not implicitly contain 
symmetry breaking corrections and quadrupole fluctuation 
correlations, and is therefore better suited for the new 
relativistic model that uses the generator coordinate method to
perform configuration mixing of angular-momentum and
particle-number projected relativistic wave functions \cite{NVR.06a,NVR.06b}.
The idea is that those correlations that we wish to treat explicitly, should 
not be included in the density functional in an implicit way, i.e. by adjusting 
parameters to data which already include correlations. The solution 
could be to adjust the functional to pseudodata, obtained by subtracting 
correlation effects from experimental masses and, eventually, radii. 
This is most easily done using masses of axially deformed nuclei with 
large deformation parameters, because the dominant contribution to 
their ground-state correlation energies is the rotational energy 
correction \cite{BBH.06}, which is relatively simple to calculate. 
Approximate methods for the calculation of correlations have been 
developed \cite{BBH.04}, that enable a systematic evaluation 
of correlation energies for the nuclear mass table. Of course one 
expects that the corresponding modifications of the parameters of the 
energy density functional will be relatively small, but even a small 
change in the relative contribution of various interaction terms could be  
the decisive factor in the description of soft potential energy surfaces, 
coexistence of prolate and oblate shapes, level ordering, etc. 
Very recent examples include the phenomenon of shape 
coexistence in neutron-deficient Kr isotopes  \cite{BBH.06a,Cle.07}, 
and the description of singular properties of excitation spectra and 
transition rates at critical points of quantum shape 
phase transitions \cite{NVLR.07}. As a first step toward the construction 
of a relativistic density functional that could provide a more accurate 
description of phenomena related to the evolution of shell structure, in this 
work we explore the possibility to determine the parameters of a given 
functional using only binding energies of axially deformed nuclei.

In Section \ref{secII} we construct the relativistic nuclear
energy density functional based on the above conjectures, and 
discuss the necessary approximations and fitting strategies. 
In Section \ref{secIII}, starting 
from microscopic nucleon self-energies in nuclear matter, and empirical 
global properties of the nuclear matter equation of state, the functional is 
accurately determined in a careful comparison of the predicted binding 
energies with data, for a set of 64 axially deformed nuclei in the 
mass regions $A\approx 150-180$ and $A\approx 230-250$. In 
Section \ref{secIV} the new energy density functional is thoroughly tested  
in a series of illustrative calculations of properties of spherical and deformed 
medium-heavy and heavy nuclei, including 
binding energies, charge radii, deformation parameters, neutron skin 
thickness, and excitation energies of giant multipole resonances. 
Section \ref{secV} summarizes the results of the present investigations and 
ends with an outlook for future studies.

\section{\label{secII} Relativistic Nuclear Energy Density Functional}
The basic building blocks of a relativistic nuclear energy density functional are the 
densities and currents
bilinear in the Dirac spinor field $\psi$ of the nucleon:
\begin{equation}
\bar{\psi}\mathcal{O}_\tau \Gamma \psi\;, \quad \mathcal{O}_\tau \in \{1,\tau_i\}\;, \quad
   \Gamma \in \{1,\gamma_\mu,\gamma_5,\gamma_5\gamma_\mu,\sigma_{\mu\nu}\}\;.
\end{equation}
Here $\tau_i$ are the isospin Pauli matrices and $\Gamma$ generically denotes the Dirac matrices. The nuclear ground-state density and energy are determined by the 
self-consistent solution of relativistic linear single-nucleon Kohn-Sham equations. 
To derive those equations it is useful to construct an 
interaction Lagrangian with four-fermion 
(contact) interaction terms in the various isospace-space channels:
\begin{center}
\begin{tabular}{ll}
 isoscalar-scalar:   &   $(\bar\psi\psi)^2$\\
 isoscalar-vector:   &   $(\bar\psi\gamma_\mu\psi)(\bar\psi\gamma^\mu\psi)$\\
 isovector-scalar:  &  $(\bar\psi\vec\tau\psi)\cdot(\bar\psi\vec\tau\psi)$\\
 isovector-vector:   &   $(\bar\psi\vec\tau\gamma_\mu\psi)
                         \cdot(\bar\psi\vec\tau\gamma^\mu\psi)$ .\\
\end{tabular}
\end{center}
Vectors in isospin space are denoted by arrows. A general Lagrangian 
can be written as a power series in the currents
 $\bar{\psi}\mathcal{O}_\tau\Gamma\psi$ and their derivatives, with higher-order 
terms representing in-medium many-body correlations
 \cite{MNH.92,Hoch.94,FML.96,RF.97,BMM.02}. 
We will adopt the approach of Refs.~\cite{FKV.04,FKV.06} and 
construct a Lagrangian with second-order 
interaction terms only, with many-body correlations encoded in 
density-dependent coupling functions. In complete analogy to the 
successful meson-exchange RMF phenomenology, in which 
the isoscalar scalar $\sigma$ meson, the isoscalar vector $\omega$ meson, 
and the isovector vector $\rho$ meson build the minimal set of meson fields that is 
necessary for a description of bulk and single-particle nuclear properties, 
we consider an effective Lagrangian that includes the isoscalar-scalar, isoscalar-vector and isovector-vector four-fermion interactions:
\begin{align}
\label{Lagrangian}
\mathcal{L} &= \bar{\psi} (i\gamma \cdot \partial -m)\psi \nonumber \\
     &- \frac{1}{2}\alpha_S(\hat{\rho})(\bar{\psi}\psi)(\bar{\psi}\psi) 
       - \frac{1}{2}\alpha_V(\hat{\rho})(\bar{\psi}\gamma^\mu\psi)(\bar{\psi}\gamma_\mu\psi) 
     - \frac{1}{2}\alpha_{TV}(\hat{\rho})(\bar{\psi}\vec{\tau}\gamma^\mu\psi)
                                                                 (\bar{\psi}\vec{\tau}\gamma_\mu\psi) \nonumber \\
    &-\frac{1}{2} \delta_S (\partial_\nu \bar{\psi}\psi)  (\partial^\nu \bar{\psi}\psi) 
         -e\bar{\psi}\gamma \cdot A \frac{(1-\tau_3)}{2}\psi\;. 
\end{align}
In addition to the free-nucleon Lagrangian and the point-coupling interaction terms, 
when applied to nuclei, the model must include the coupling of the protons to the
electromagnetic field. 
The derivative term in Eq.~(\ref{Lagrangian}) accounts for leading effects of finite-range
interactions that are crucial for a quantitative description of nuclear density distribution, e.g. nuclear radii. Similar interactions can be included in each space-isospace channel, but in practice data on charge radii constrain only a single derivative term,
for instance $\delta_S (\partial_\nu \bar{\psi}\psi)  (\partial^\nu \bar{\psi}\psi) $.  
The coupling parameter $\delta_S$ has been estimated, for instance, 
from in-medium chiral perturbation calculation of inhomogeneous nuclear 
matter~\cite{FKW.05}. In the region
of nucleon densities relevant for the description of finite nuclei 
($0.1~\textnormal{fm}^{-3} \le \rho \le  0.15~\textnormal{fm}^{-3}$), 
the coupling strength of the derivative term displays a rather weak density 
dependence and can be approximated by a constant value $\delta_S$  between
$-0.85\; \textnormal{fm}^4$ and $-0.7\; \textnormal{fm}^4$.
Note that the inclusion of an adjustable derivative term only
in the isoscalar-scalar channel is consistent with conventional meson-exchange 
RMF models, in which the mass of the fictitious $\sigma$ meson is adjusted to
nuclear matter and
ground-state properties of finite nuclei, whereas free values are used for the 
masses of the $\omega$ and $\rho$ mesons. 

The point-coupling Lagrangian Eq.~(\ref{Lagrangian}) does not 
include isovector-scalar terms. In the meson-exchange
picture this channel is represented by the exchange of an effective $\delta$ meson, 
and its inclusion introduces a proton-neutron effective mass splitting and enhances 
the isovector spin-orbit potential. 
However, in calculations of ground-state properties of finite nuclei, using
both meson-exchange \cite{TW.99,HKL.01} and point-coupling \cite{BMM.02}
models, it has not been possible to constrain the parameters of the effective 
interaction in the isovector-scalar channel. Although the isovector strength has 
a relatively well-defined value, the distribution between the scalar and vector 
channels is not determined by ground-state data. To reduce the number of 
adjustable parameters, the isovector-scalar channel may be omitted from
an energy density functional that will primarily be used for the description of 
low-energy nuclear structure.

The strength parameters of the interaction terms in Eq. (\ref{Lagrangian}) 
are functions of the nucleon 4-current:
\begin{equation}
j^\mu = \bar{\psi} \gamma^\mu \psi = \hat{\rho} u^{\mu} \; ,
\end{equation}
where $u^{\mu}$ is the 4-velocity defined as 
$(1-{\bm v}^2)^{-1/2}(1,{\bm v})$. In the rest-frame of the 
nuclear system: $\bm v=0$. The single-nucleon 
Dirac equation, the relativistic analogue of the Kohn-Sham equation, is obtained
from the variation of the Lagrangian with respect to $\bar{\psi}$:
\begin{equation}
\left[ \gamma_\mu(i\partial^\mu - \Sigma^\mu -\Sigma_R^\mu) - (m+\Sigma_S)\right]\psi = 0\;,
\label{Dirac-eq}
\end{equation}
with the nucleon self-energies defined by the following relations:
\begin{align}
\label{sigma_v}
\Sigma^\mu &= \alpha_V(\rho_v) j^\mu + e  \frac{(1-\tau_3)}{2} A^\mu\\
\label{sigma_r}
\Sigma_R^\mu &= \frac{1}{2}\frac{j^\mu}{\rho_v}
            \left\{ \frac{\partial \alpha_S}{\partial \rho}\rho_s^2
         +\frac{\partial \alpha_V}{\partial \rho}j_\mu j^\mu 
         + \frac{\partial \alpha_{TV}}{\partial \rho}\vec{j}_\mu \vec{j}^\mu
          \right\}\\
\label{sigma_s}
\Sigma_S &= \alpha_S(\rho_v)\rho_s - \delta_S \Box \rho_s \\
\label{sigma_tv}
\Sigma_{TV}^\mu &=  \alpha_{TV}(\rho_v)\vec{j}^\mu \;.
\end{align}
In addition to the contributions of the isoscalar-vector four-fermion interaction and 
the electromagnetic interaction, the isoscalar-vector self-energy $\Sigma^\mu$ 
includes the ``rearrangement'' 
terms $\Sigma_R^\mu$, arising from the variation of the vertex 
functionals $\alpha_S$, $\alpha_V$, and $\alpha_{TV}$ with respect to the 
nucleon fields in the density operator $\hat{\rho}$. 
The inclusion of the rearrangement self-energy is essential for 
energy-momentum conservation and the thermodynamical consistency of 
the model \cite{FLW.95,TW.99,NVFR.02}. $\Sigma_S$  and $\Sigma_{TV}^\mu$
denote the isoscalar-scalar and isovector-vector self-energies, respectively.

In the relativistic density functional framework the nuclear ground state $\rightg$ is 
represented by the mean-field self-consistent solution of the system of equations 
(\ref{Dirac-eq}) -- (\ref{sigma_tv}), with the 
isoscalar and isovector 4-currents and 
scalar density:
\begin{eqnarray}
\label{den1}
j_\mu & = \leftg \bar{\psi} \gamma_\mu \psi \rightg =
& \sum_{k=1}^N v_{k}^{2}~\bar{\psi}_k \gamma_\mu \psi_k \; ,\\
\label{den2}
\vec{j}_\mu & =  
\leftg \bar{\psi} \gamma_\mu \vec{\tau} \psi \rightg =
& \sum_{k=1}^N v_{k}^{2}~\bar{\psi}_k \gamma_\mu \vec{\tau} \psi_k \; ,\\
\label{den3}
\rho_S & = \leftg \bar{\psi} \psi \rightg = 
& \sum_{k=1}^N v_{k}^{2}~\bar{\psi}_k \psi_k \; ,
\end{eqnarray}
where $\psi_k$ are Dirac spinors, and
the sum runs over occupied positive-energy single-nucleon orbitals,
including the corresponding occupation factors $v_{k}^{2}$.
The single-nucleon Dirac equations are solved
self-consistently in the ``no-sea''
approximation that omits the explicit contribution 
of negative-energy solutions of the relativistic 
equations to the densities and currents. Vacuum polarization effects 
are implicitly included in the adjustable density-dependent parameters of the theory.

A large part of this work will be devoted to adjusting the free parameters of the 
medium-dependent point-coupling functionals 
$\alpha_S$, $\alpha_V$, and $\alpha_{TV}$,  
and the strength $\delta_S$ of the derivative term. To establish the density 
dependence of the couplings one 
could start from a microscopic (relativistic) EoS of symmetric 
and asymmetric nuclear matter, and map the corresponding nucleon self-energies 
on the mean-field self-energies Eqs.~(\ref{sigma_v}) - (\ref{sigma_tv}) that 
determine the single-nucleon Dirac equation (\ref{Dirac-eq}). This approach has 
been adopted, for instance, in RMF models based on Dirac-Brueckner-Hartree-Fock 
self-energies in nuclear matter \cite{FLW.95,JL.98,HKL.01}, or on in-medium 
chiral perturbation theory (ChPT) calculations of the nuclear matter 
EoS \cite{FKV.04,FKV.06}. In general, however, energy density functionals determined 
directly from a microscopic EoS do not provide a very accurate description of data 
in finite nuclei. The reason, of course, is that a calculation of the nuclear matter EoS 
involves approximation schemes and includes adjustable parameters that are not 
really constrained by nuclear structure data. The resulting bulk properties of 
infinite nuclear matter (saturation density, binding energy, compression modulus, 
asymmetry energy) do not determine uniquely the parameters of nuclear 
energy density functionals, which usually must be further fine-tuned  
to ground-state data (masses and/or charge radii) of spherical nuclei.

In a phenomenological construction of a relativistic energy density functional 
one starts from an assumed ansatz for the medium dependence of the mean-field 
nucleon self-energies, and adjusts the parameters directly adjusted to data of 
spherical nuclei. This procedure 
was used, for instance, in the construction of the relativistic density-dependent 
interactions TW-99~\cite{TW.99}, DD-ME1~\cite{NVFR.02}, 
DD-ME2~\cite{LNVR.05}, PKDD~\cite{LMGZ.04}, PK01~\cite{Long.06}.

This work adopts a different strategy and determines the parameters 
of the point-coupling Lagrangian Eq.~(\ref{Lagrangian}) exclusively 
from a large data set of 
binding energies of deformed nuclei. First one notes that calculated masses of finite 
nuclei are primarily sensitive to the three leading terms in the empirical mass formula: volume, surface and symmetry energy
\begin{equation}
\label{SEMF}
B.E. = a_v A +a_s A^{2/3}+a_4\frac{(N-Z)^2}{4A}+\cdots\;.
\end{equation}
Therefore one can generate families of effective interactions that are characterized 
by different values of $a_v$, $a_s$ and $a_4$, and determine which parametrization 
minimizes the deviation from the empirical binding energies of a large set of 
deformed nuclei. This approach differs considerably from the standard procedure of 
fitting parameters of nonrelativistic Skyrme or RMF functionals, in which a given 
set of parameters is adjusted simultaneously to a favorite nuclear matter EoS and 
to ground-state properties of about 10 spherical closed-shell nuclei.
Deformed systems have 
generally not been included in fits of parameters of self-consistent RMF models. 
The reason, of course, is that calculation of deformed nuclei is computationally 
more demanding and requires advanced computer codes. In this work 
parameters of relativistic energy density functionals are for the first time 
directly adjusted to binding energies of axially deformed nuclei in the 
mass regions $A\approx 150-180$ and $A\approx 230-250$. 
 
To determine the functional form of the density dependence of the couplings 
$\alpha_S$, $\alpha_V$, and $\alpha_{TV}$, one can start from microscopic 
nucleon self-energies in nuclear matter. In a recent analysis of relativistic 
nuclear dynamics \cite{PF.06}, modern high-precision nucleon-nucleon (NN) 
potentials (Argonne V$_{18}$, Bonn A, CD-Bonn, Idaho, Nijmegen, V$_{low k}$) 
were mapped on a relativistic operator basis, and the corresponding relativistic 
nucleon self-energies in nuclear matter were calculated in Hartree-Fock 
approximation at {\em tree level}. A very interesting result is that, at moderate 
nucleon densities relevant for nuclear structure calculations, all potentials 
yield very similar scalar and vector mean fields of several hundred MeV 
magnitude, in remarkable agreement 
with standard RMF phenomenology: at saturation density a large and attractive 
scalar field $\Sigma_s \approx -400$ MeV, and a repulsive vector field 
$\Sigma_v \approx 350$ MeV. The different treatment of short-distance 
dynamics in the various NN potentials leads to slightly more pronounced 
differences between the corresponding self-energies at higher nucleon densities. 
Generally, however, all potentials predict a very 
similar density dependence of the scalar and vector self-energies. 
In the chiral effective field theory framework, in particular, these self-energies are 
predominantly generated by contact terms that occur at next-to-leading order in 
the chiral expansion. 

Of course at the Hartree-Fock {\em tree level} these NN potentials do not yield 
saturation of nuclear matter. Nevertheless, the corresponding self-energies can 
be used as the starting point in the modeling of medium dependence of a 
relativistic nuclear energy density functional. 
Guided by the microscopic density dependence of the 
vector and scalar self-energies, we choose the following practical ansatz for 
the functional form of the couplings 
\begin{equation}
\alpha_i(\rho) = a_i + (b_i+c_i x)e^{-d_ix}\quad\quad (i\equiv S, V, TV)\;, 
\label{ansatz}
\end{equation}
with $x=\rho/\rho_{sat}$, and $\rho_{sat}$ denotes the nucleon density at saturation in symmetric nuclear matter. Note that the corresponding self-energies are defined in 
Eqs.~(\ref{sigma_v}) - (\ref{sigma_tv}). In the next section we will adjust the 
parameters of the ansatz  Eq.~(\ref{ansatz}) simultaneously to infinite and semi-infinite
nuclear matter, and to binding energies of deformed nuclei. The resulting self-energies 
in nuclear matter will eventually be compared to our starting approximation: 
the Hartree-Fock scalar and vector self-energies of the Idaho N$^3$LO 
potential \cite{EM.03}.

In the isovector channel the corresponding Hartree-Fock {\em tree level} nucleon 
self-energies, obtained by directly mapping microscopic NN potentials on a 
relativistic operator basis, are presently not available. Therefore, as it was done 
in the case of the finite-range meson exchange interactions 
TW-99 \cite{TW.99}, DD-ME1 \cite{NVFR.02}, DD-ME2 \cite{LNVR.05}, and
PK01 \cite{Long.06}, the density dependence of the isovector-vector coupling
function is determined from the results of Dirac-Brueckner calculations of asymmetric 
nuclear matter \cite{JL.98}. Accordingly, in Eq.~(\ref{ansatz}) for the 
isovector channel we set two parameters to zero: $a_{TV}=0$ and $c_{TV}=0$, 
and adjust $b_{TV}$ and $d_{TV}$ to 
empirical properties of asymmetric matter and to nuclear masses, together 
with the parameters of the isoscalar channel.
\section{\label{secIII} The effective density-dependent interaction DD-PC1}
\subsection{\label{secIIIa}Infinite and semi-infinite nuclear matter}

The usual procedure in the construction of an effective mean-field 
interaction is the {\em least-squares} adjustment of 
parameters to both nuclear matter EoS and to ground-state data (masses, 
charge radii) of spherical nuclei. Instead we generate sets of effective 
interactions with different values of the volume $a_v$, surface $a_s$, and 
symmetry energy $a_4$ in nuclear matter, and analyze the corresponding binding 
energies of deformed nuclei with $A\approx 150-180$ and $A\approx 230-250$. 
The nuclear matter saturation density, compression modulus, 
and Dirac mass will be kept fixed throughout this analysis. The calculated 
binding energies of finite nuclei are not very sensitive to the
nuclear matter saturation density, and we take $\rho_{sat}=0.152~\textnormal{fm}^{-3}$,
in accordance with values predicted by most modern relativistic mean-field models. 
In particular, this value has also been used for the meson-exchange effective  
interactions DD-ME1 \cite{NVFR.02}, and DD-ME2 \cite {LNVR.05}. From these 
interactions we also take the Dirac effective nucleon mass 
$m^*_D =  m + \Sigma_S = 0.58 m$. In RMF theory the Dirac mass is closely 
related to the effective spin-orbit single-nucleon potential, and empirical 
energy spacings between spin-orbit partner states in finite nuclei determine 
a relatively narrow interval of allowed values: $0.57 \le m^*_D/m \le 0.61$. 
In a recent study \cite{NVLR.08} of the relation between finite-range 
(meson-exchange) and zero-range (point-coupling) representations of 
effective RMF interactions we have shown that,  
to reproduce experimental excitation energies of isoscalar giant
monopole resonances, point-coupling interactions require a
nuclear matter compression modulus $K_{nm}\approx 230$ MeV, 
considerably lower than values typically used for finite-range
meson-exchange relativistic interactions. Thus we take 
$K_{nm} = 230$ MeV for all effective interactions 
considered in the present analysis. 

Of course if only nuclear matter properties at the point of saturation 
density were specified, one could parametrize a number of realistic 
effective interactions that would be difficult to compare at the level of 
finite nuclei. In particular, nuclear structure data do not constrain the 
nuclear matter EoS at high nucleon densities. Therefore, in addition to 
$\rho_{sat}$, $m^*_D$, and $K_{nm}$, we fix two additional points on the 
$E(\rho)$ curve in symmetric matter to the microscopic EoS of Akmal, 
Pandharipande and Ravenhall \cite{APR.98}, based on the Argonne V$_{18}$ 
NN potential and the UIX three-nucleon interaction. This EoS has extensively been 
used in studies of high-density nucleon matter and neutron stars. At almost 
four times nuclear matter saturation density, 
we choose the point $\rho =0.56~\textnormal{fm}^{-3}$ with $E/A = 34.39$ MeV and, 
to have an overall consistency, one point at low density:  
$\rho =0.04~\textnormal{fm}^{-3}$ with $E/A = -6.48$ MeV
(cf. Table VI of Ref.~\cite{APR.98}). As we have already emphasized in the 
previous section, by adjusting mean-field  
interactions exclusively to a microscopic EoS like, for instance, the one calculated 
in Ref.~\cite{APR.98}, it is not possible to obtain a very accurate description of nuclear
structure. Ground-state nuclear data must be used to fine tune the parameters 
of effective interactions.

\begin{figure}
\includegraphics[scale=0.7,angle=270]{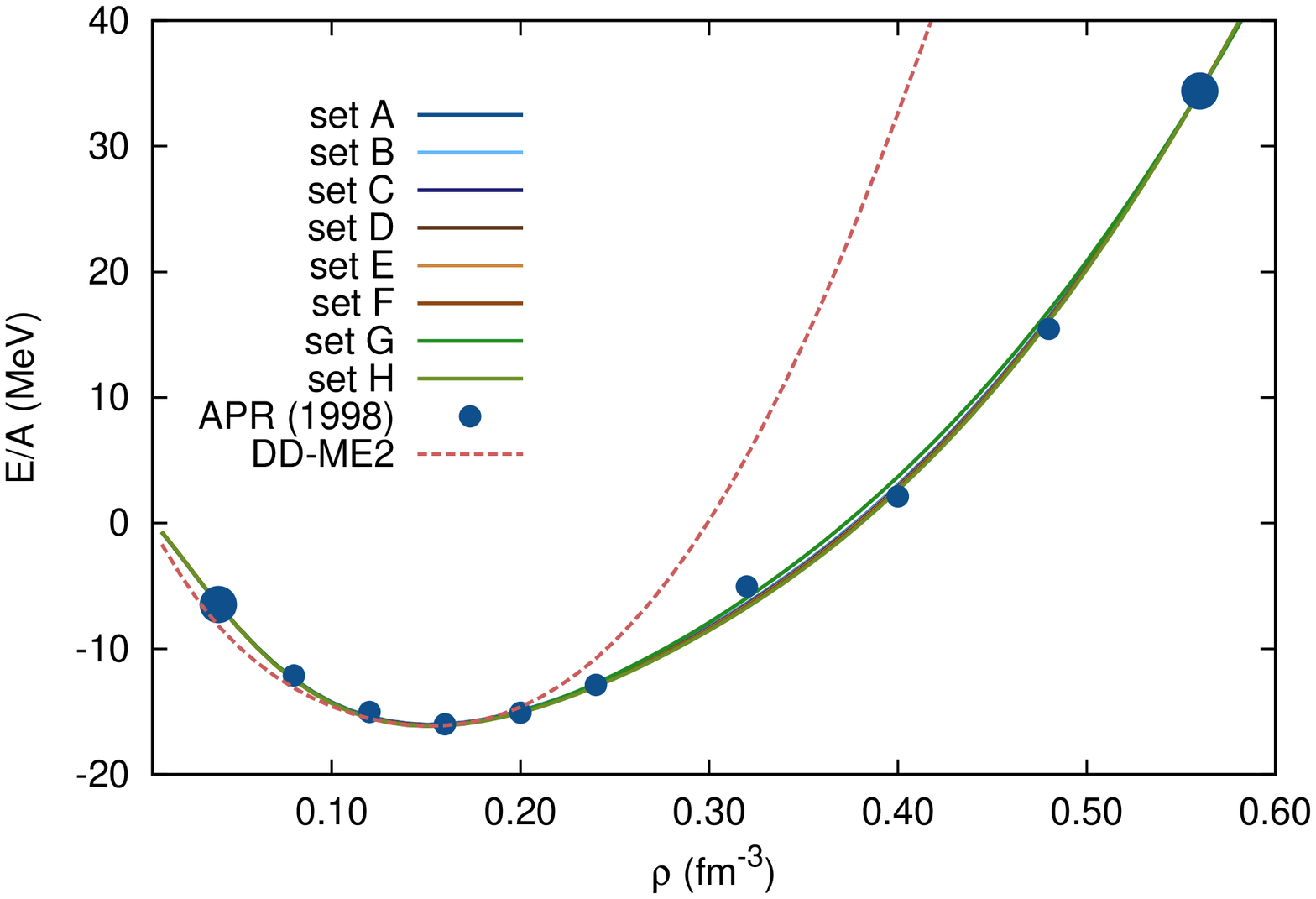}
\vspace{1cm}
\caption{(Color online) 
The equations of state of symmetric nuclear matter (binding energy
as a function of nucleon density) for the eight point-coupling effective interactions
of Table \ref{TabC}, in comparison with the EoS of the meson-exchange effective interaction DD-ME2 \cite{LNVR.05}, and the microscopic EoS of Ref.~\cite{APR.98}. 
The two points from the microscopic EoS on which the point-coupling effective 
interactions (sets A--H) were adjusted, are denoted by larger filled circle symbols.}
\label{FigA}
\end{figure}

In contrast to the Dirac mass and saturation density,  the nuclear matter volume 
energy coefficient $a_v$ has a decisive influence on the calculated binding energies 
of finite nuclei. Using the framework of nonrelativistic Skyrme functionals, 
it was recently shown that even a relatively small change in the volume energy 
($\approx 0.5\%$) can have a pronounced effect on the calculated masses of 
heavy and superheavy nuclei, as compared with experimental 
values \cite{BBH.06,BSU.05}.  In the framework of RMF models 
no attempt has been made so far to constrain the value of volume
energy better than the interval 
$-16.2~ \textnormal{MeV} \le a_v \le -16~ \textnormal{MeV}$.
To study in more detail the effect of volume energy on masses,
we have generated point-coupling effective interactions characterized by
the following values of the coefficient: 
$a_v=-16.02$ MeV (set A), $a_v=-16.04$ MeV (set B), $a_v=-16.06$ MeV (set C),
$a_v=-16.08$ MeV (set D) , $a_v=-16.10$ MeV (set E), $a_v=-16.12$ MeV (set F),
$a_v=-16.14$ MeV (set G) and $a_v=-16.16$ MeV (set H). 
The corresponding parameters of the ansatz Eq.~(\ref{ansatz}) for 
the functional form of the isoscalar couplings, are collected in Table \ref{TabC}. 
Note that to reduce the number of free parameters, we have set the value $c_V=0$.
The resulting binding energy curves for symmetric nuclear matter are plotted in 
Fig.~\ref{FigA}, together with the EoS of the meson-exchange effective interaction 
DD-ME2 , and the microscopic EoS of Ref.~\cite{APR.98}. The two points on 
the microscopic EoS that have been used to adjust the parameters are 
represented by large filled circle symbols. Because of the anchor at 
$\rho =0.56~\textnormal{fm}^{-3}$, the new binding energy curves are,
of course, different from DD-ME2 and 
much closer to the microscopic EoS. However, the high-density behavior does not 
influence much the description of low-energy nuclear structure data.

The isovector channel of the energy density functional determines the 
density dependence of the nuclear matter symmetry energy
\begin{equation}
\label{S2}
S_2(\rho) = a_4+\frac{p_0}{\rho_{sat}^2}(\rho-\rho_{sat}) 
                         + \frac{\Delta K_0}{18 \rho_{sat}^2}(\rho-\rho_{sat})^2+\cdots\;.
\end{equation}
The parameter $p_0$ characterizes the linear density dependence of the 
symmetry energy, and $\Delta K_0$ is the isovector correction to 
the compression modulus.  Experimental masses, unfortunately, do not 
place very strict constraints on the parameters of the expansion 
of $S_2(\rho)$ \cite{Fur.02}, but self-consistent mean-field calculations
show that binding energies can restrict the values of 
$S_2$ at nucleon densities somewhat 
below saturation density, i.e. at $ \rho  \approx 0.1~\textnormal{fm}^{-3}$.
Additional information on the symmetry energy can be obtained from 
data on neutron skin thickness and excitation energies of giant dipole 
resonances. Although values of neutron radii are available only for a 
small number of nuclei and the corresponding uncertainties are
large, recent studies have shown that relativistic effective interactions
with volume asymmetry $a_4$ in the range 
$31\;\textnormal{MeV} \le a_4 \le 35\; \textnormal{MeV}$ 
predict values for neutron skin thickness that are consistent with data, 
and reproduce experimental excitation 
energies of isovector giant dipole resonances
(cf. Ref.~\cite{VNR.03} and references therein cited). 
Therefore we keep the volume asymmetry fixed at $a_4=33\;MeV$, 
and vary the symmetry energy at a density that corresponds to 
an average nucleon density in finite nuclei:
$\langle \rho \rangle =0.12\;\textnormal{fm}^{-3}$. 
The quantity $S_2(\rho=0.12\;\textnormal{fm}^{-3})$ will be 
denoted $\langle S_2 \rangle$. 

\begin{figure}
\includegraphics[scale=0.7,angle=270]{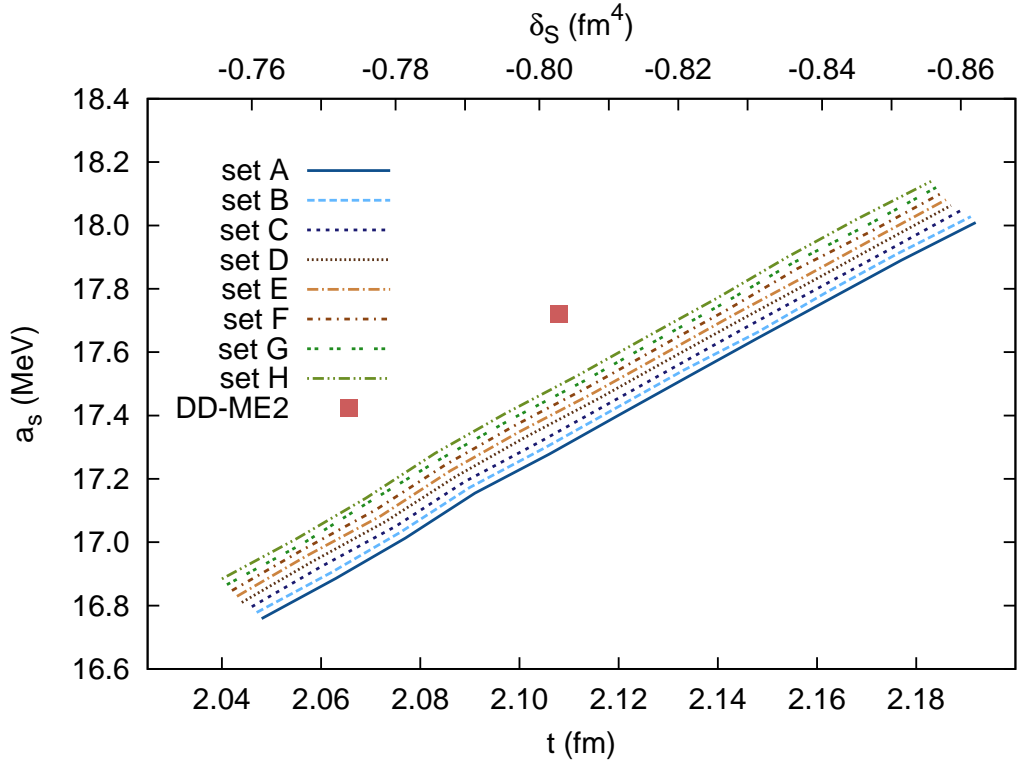}
\vspace{1cm}
\caption{(Color online)
Surface energy of semi-infinite nuclear matter as a function
of the surface thickness, for the eight point-coupling effective interactions 
of Table \ref{TabC}. The corresponding values of the 
strength $\delta_S$ of the derivative coupling 
term in the point-coupling Lagrangian Eq.~(\ref{Lagrangian}), are 
displayed on the upper horizontal axis. The filled square symbol 
denotes the surface energy and surface thickness predicted by the 
meson-exchange effective interaction DD-ME2 \cite{LNVR.05}.
}
\label{FigB}
\end{figure}

Calculated binding energies and charge radii are strongly influenced by the 
choice of the surface energy coefficient $a_s$. In the present model the value 
of this quantity is determined by the strength $\delta_S$ of the derivative coupling 
term in the point-coupling Lagrangian Eq.~(\ref{Lagrangian}). 
For each effective interaction (sets A-H), we have calculated
the surface energy and surface thickness of semi-infinite nuclear matter \cite{HF.89}, 
for several values of the parameter $\delta_s$ in the range  
$-0.76\; \textnormal{fm}^4 \le \delta_S \le -0.86\; \textnormal{fm}^4$.
In Fig.~\ref{FigB} we plot the corresponding surface energies as  
functions of the surface thickness, in comparison to the point 
obtained with the finite-range interaction DD-ME2 ($t=2.108$ fm, $a_s=17.72$ MeV). 
Considering that DD-ME2 has an {\em rms} error of only 0.017 fm 
when compared to data on absolute charge radii and charge isotope 
shifts \cite{LNVR.05}, and also taking into account 
the comparison between DD-ME2 and point-coupling RMF 
interactions of Ref.~\cite{NVLR.08}, the following range for 
the parameter of the derivative coupling term can be deduced:
$-0.80\; \textnormal{fm}^4 \ge \delta_S \ge -0.84\; \textnormal{fm}^4$, 
in very good agreement with the microscopic estimate of Ref.~\cite{FKW.05} 
for the region of nucleon densities $\rho \approx 0.1 \; \textnormal{fm}^3$.

\subsection{\label{secIIIb}Deformed nuclei }
If an effective interaction is adjusted to masses of finite nuclei 
by varying the volume, symmetry, and surface energies, the parameters 
of the energy density functional that determine 
these quantities will generally be correlated because of 
Eq.~(\ref {SEMF}). When only a small number of nuclei is considered, satisfactory 
results can be obtained with various, in general linearly dependent 
combinations of parameters. The new effective 
point-coupling interactions will therefore be analyzed on a set of $64$ deformed nuclei,
listed in Table~\ref{TabD}. To resolve the surface and volume contributions to 
binding energy, nuclides with mass number ranging from $154$ to
$250$ are considered. The variation of  the asymmetry coefficient
\begin{equation}
\label{alpha}
\alpha^2 = \frac{(N-Z)^2}{A^2}
\end{equation}
in the range from $0.018$ to $0.054$, should suffice to deduce the isovector 
parameters that govern the symmetry energy contribution.  
The effect of shell closure is minimized by taking into account only well deformed nuclei. 
Pairing correlations are treated in the BCS approximation with empirical pairing 
gaps (5-point formula). The pairing model space includes two major oscillator 
shells ($2\hbar\omega_0$) above the Fermi surface. 
The self-consistent single-nucleon RMF equations are solved 
by expanding nucleon spinors in terms of eigenfunctions of a deformed, 
axially symmetric harmonic oscillator potential \cite{CPC}. In this work calculations 
of nuclear ground states are performed in a large basis of 16 major oscillator 
shells, and convergence has been tested in several calculations 
with 18 oscillator shells. After the self-consistent
equations are solved, the microscopic estimate for the center-of-mass 
correction is subtracted from the total binding energy 
\begin{equation}
\label{cms}
E_{cm} = - \frac{<P_{cm}^2>}{2Am} \;,
\end{equation}
where $P_{cm}$ is the total momentum of a nucleus with $A$ nucleons.

\begin{figure}
\includegraphics[scale=0.7,angle=270]{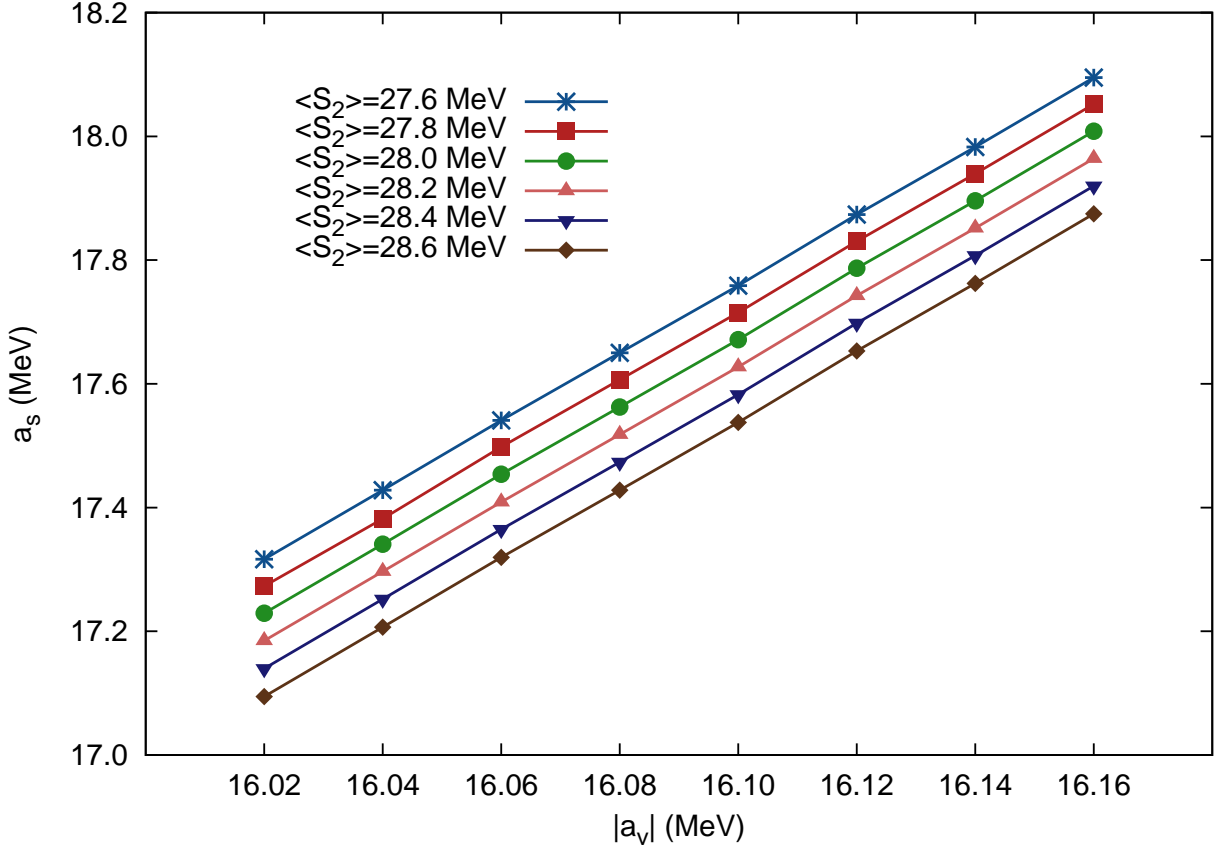}
\vspace{2cm}
\caption{(Color online)
Surface energies of semi-infinite nuclear matter that minimize 
the deviation of the calculated binding
energies from data, for the set of nuclei listed in Table \ref{TabD}, plotted  
as functions of the volume energy at saturation, for six values of 
the symmetry energy $\langle S_2 \rangle$.
}
\label{FigC}
\end{figure}

\begin{figure}
\includegraphics[scale=0.7,angle=270]{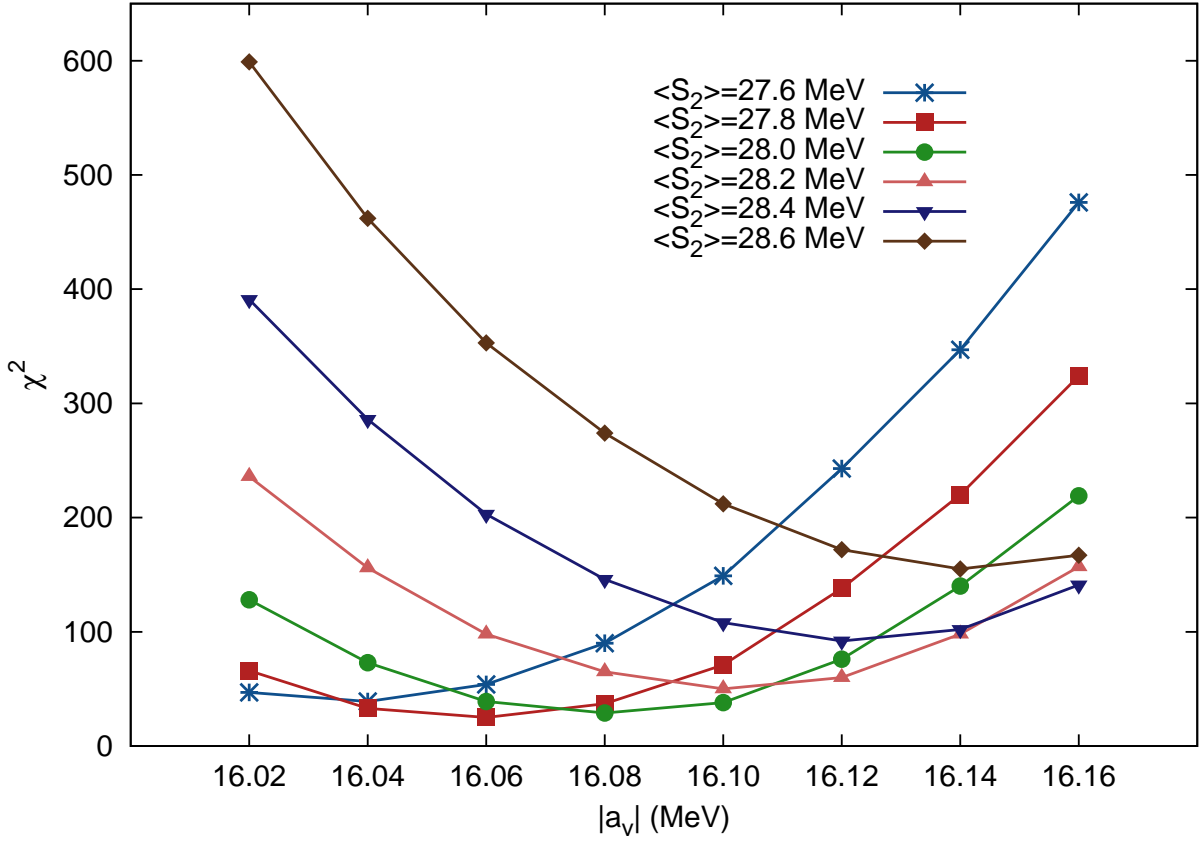}
\vspace{2cm}
\caption{(Color online) 
$\chi^2$-deviations Eq.~(\ref{chi2}) of the theoretical  binding
energies from data for the set of deformed nuclei listed
in Table~\ref{TabD}, and for each combination of the surface, volume and symmetry
energy shown in Fig.~\ref{FigC}.
}
\label{FigD}
\end{figure}

\begin{figure}
\includegraphics[scale=0.7,angle=270]{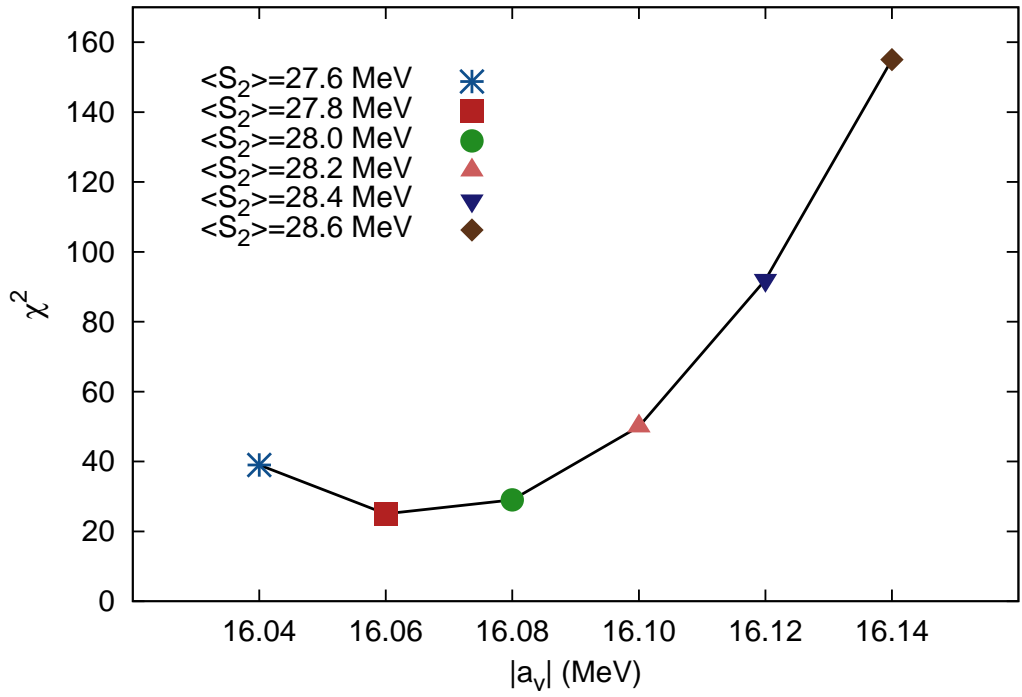}
\vspace{1cm}
\caption{(Color online)
The minimum $\chi^2$-deviation of the theoretical  binding
energies from data, as a function of the volume energy coefficient.
Each point represents the minimum of the corresponding curve plotted 
in Fig.~\ref{FigD}.
}
\label{FigE}
\end{figure}

For each effective interaction with given volume energy $a_v$ 
(sets A-H), and for six values of the symmetry
energy $\langle S_2 \rangle = 27.6$, 27.8, 28.0, 28.2, 28.4, and 28.6 MeV,  we have adjusted the surface energy (i.e. the coupling strength $\delta_S$ of
the derivative term in the Lagrangian Eq.~(\ref{Lagrangian})) to a value that 
minimizes the deviation of the calculated binding
energies from data, for the set of nuclei listed in Table \ref{TabD}. 
The required accuracy is 0.05\%, which approximately corresponds to an
absolute error of $\pm 1$ MeV for the total binding energy.
The resulting surface energies are plotted in Fig.~\ref{FigC} as functions of the
volume energy, for each value of the symmetry energy $\langle S_2 \rangle$.
At this point we have a set of $48$ parametrizations of the energy 
density functional. Fig.~\ref{FigD} displays the corresponding $\chi^2$-values 
\begin{equation}
\chi^2 = \sum_i \left ( {{E_B^{th}(i) - E_B^{exp}(i)}\over \Delta E_B^{exp}(i)} \right )^2 \; ,
\label{chi2}
\end{equation}
where $E_B^{exp}(i)$ denote experimental binding energies \cite{AW.03}, 
$E_B^{th}(i)$ are the corresponding theoretical values, 
$\Delta E_B^{exp}(i) = 0.0005 E_B^{exp}(i)$, and the sum runs over 
the set of 64 deformed nuclei. Although the span of $\chi^2$-values is very large, 
the functional dependence of  $\chi^2$ on $a_v$ is smooth and, 
for each value $\langle S_2 \rangle$ 
of the symmetry energy, there is a unique 
combination of volume and surface energies that minimizes 
$\chi^2$. The minima of each curve are collected in Fig.~\ref{FigE}. Also
in this plot $\chi^2$ {\em vs} $a_v$ displays a smooth parabola, with 
the absolute minimum at the point $a_v=-16.06$ MeV,  
$\langle S_2 \rangle=27.8$ MeV, and $a_s=17.498$ MeV.  The    
$\chi^2$-values of the neighboring points are not much larger, but obviously 
the systematics of binding energies excludes effective interactions 
with $a_v \leq -16.10$ MeV. 

\begin{figure}
\includegraphics[scale=0.7]{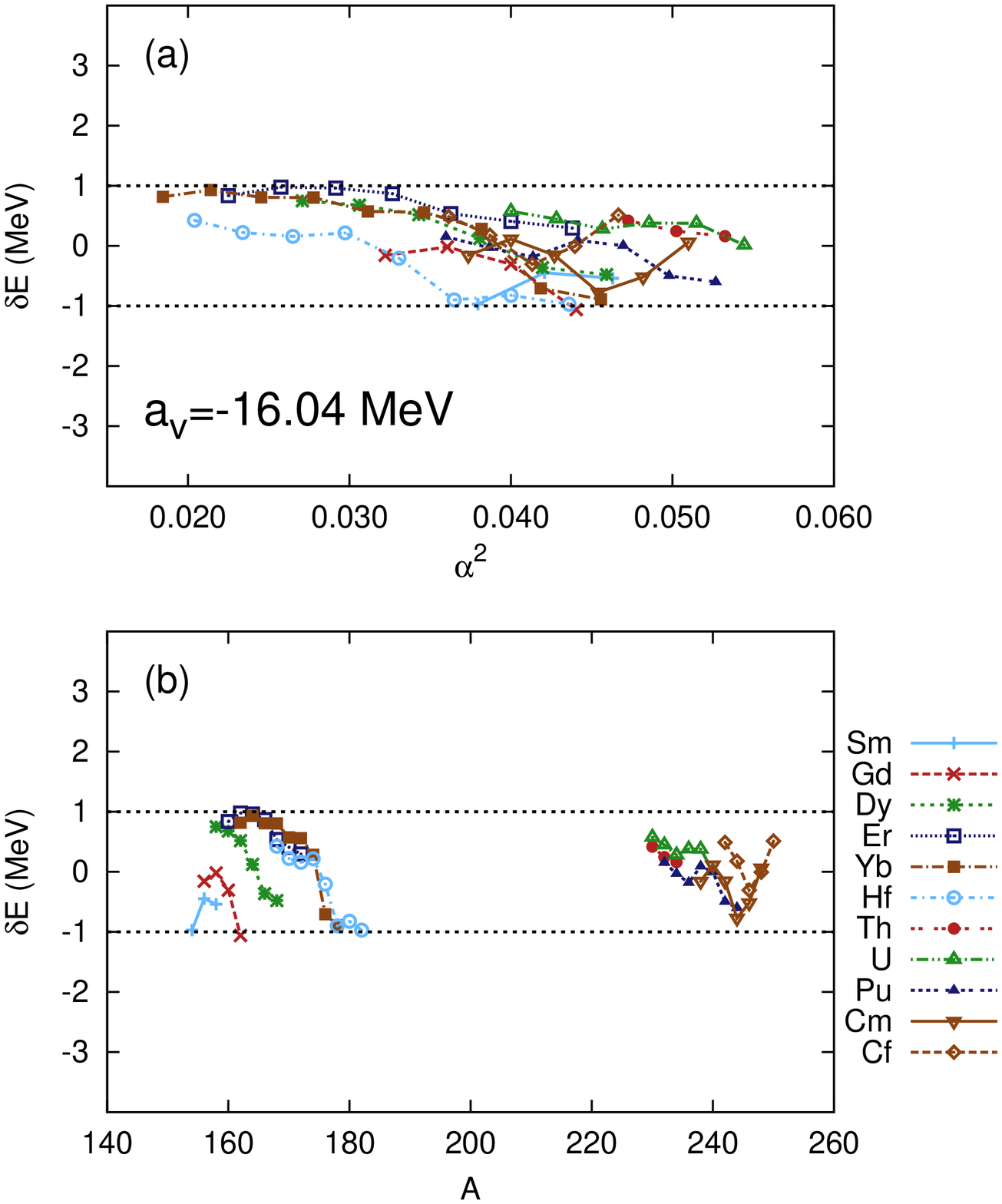}
\caption{(Color online)
Absolute deviations of the calculated binding energies from the experimental 
values of the 64 axially deformed nuclei listed in Table \ref{TabD}, 
as functions of the asymmetry coefficient (upper panel), and 
mass number (lower panel). Lines connect nuclei that 
belong to the isotopic chains shown in the legend. 
The theoretical binding energies are calculated using the point-coupling
effective interaction characterized by the volume energy $a_v=-16.04$ MeV.}
\label{FigF}
\end{figure}

\begin{figure}
\includegraphics[scale=0.7]{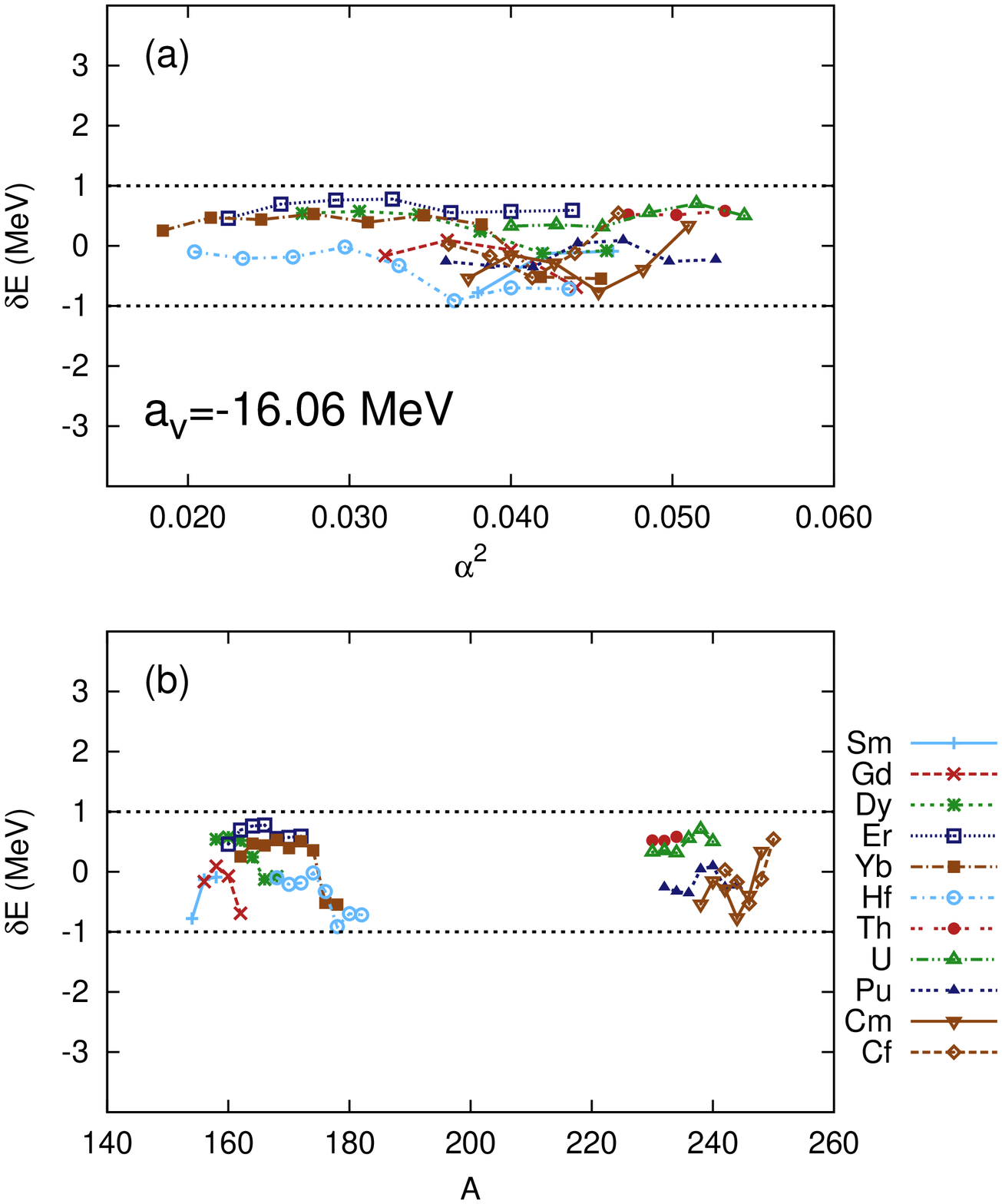}
\caption{(Color online)
Same as in Fig.~\ref{FigF}, for the point-coupling effective interaction with volume
energy $a_v=-16.06$ MeV.
}
\label{FigG}
\end{figure}

\begin{figure}
\includegraphics[scale=0.7]{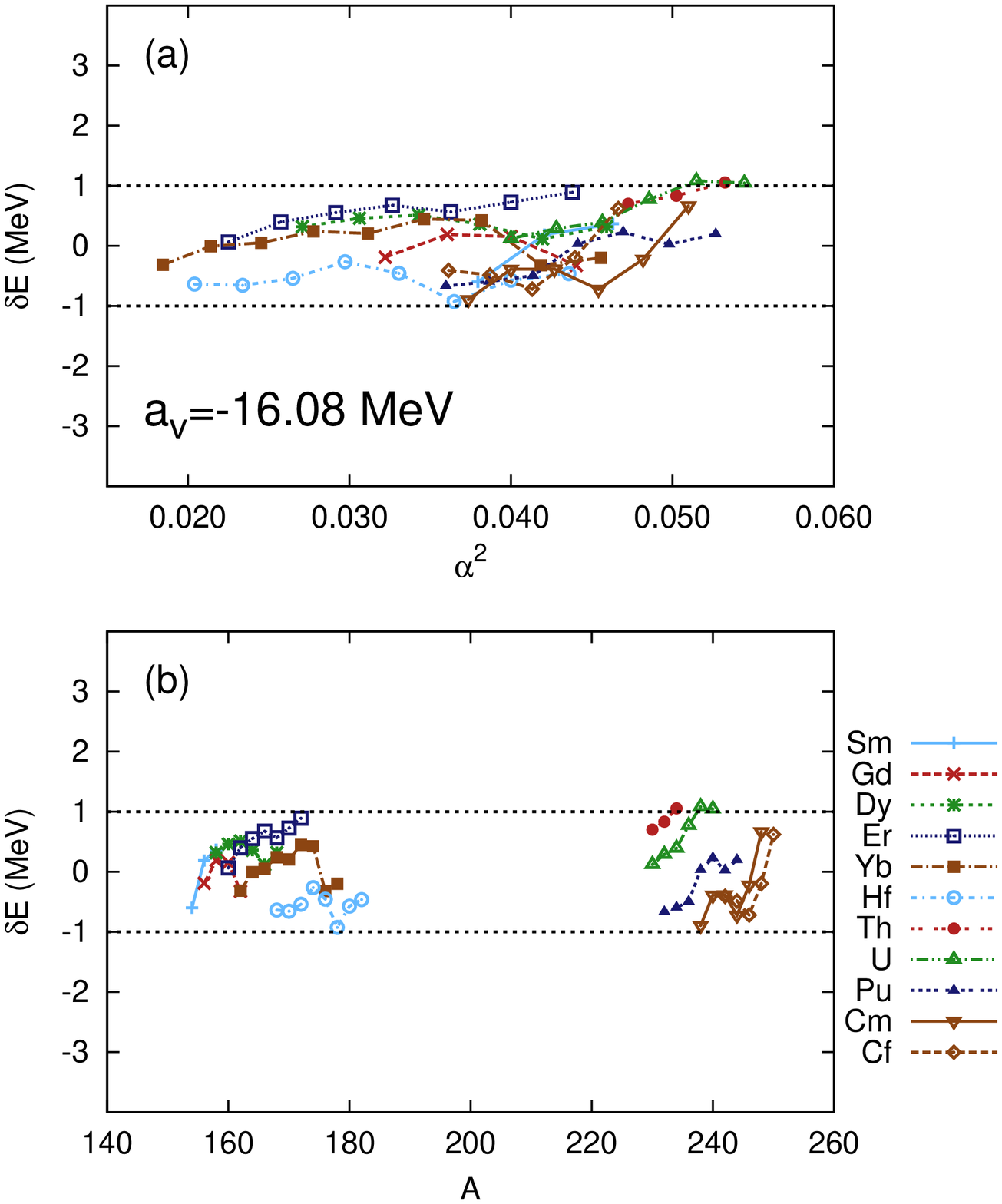}
\caption{(Color online)
Same as in Fig.~\ref{FigF}, for the point-coupling effective interaction with volume
energy $a_v=-16.08$ MeV.
}
\label{FigH}
\end{figure}

\begin{figure}
\includegraphics[scale=0.7]{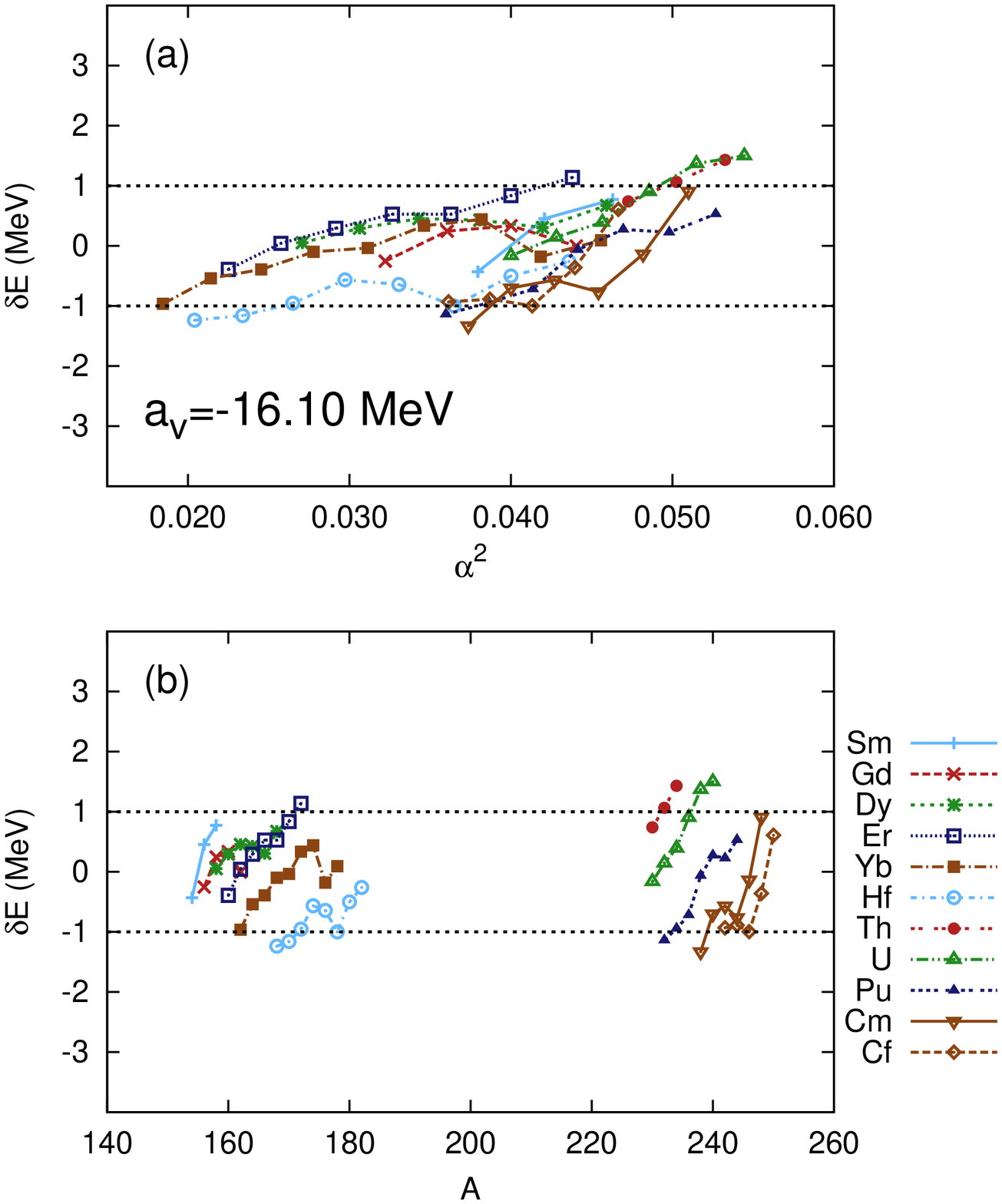}
\caption{(Color online)
Same as in Fig.~\ref{FigF}, for the point-coupling effective interaction with volume
energy $a_v=-16.10$ MeV.
}
\label{FigI}
\end{figure}

\begin{figure}
\includegraphics[scale=0.7]{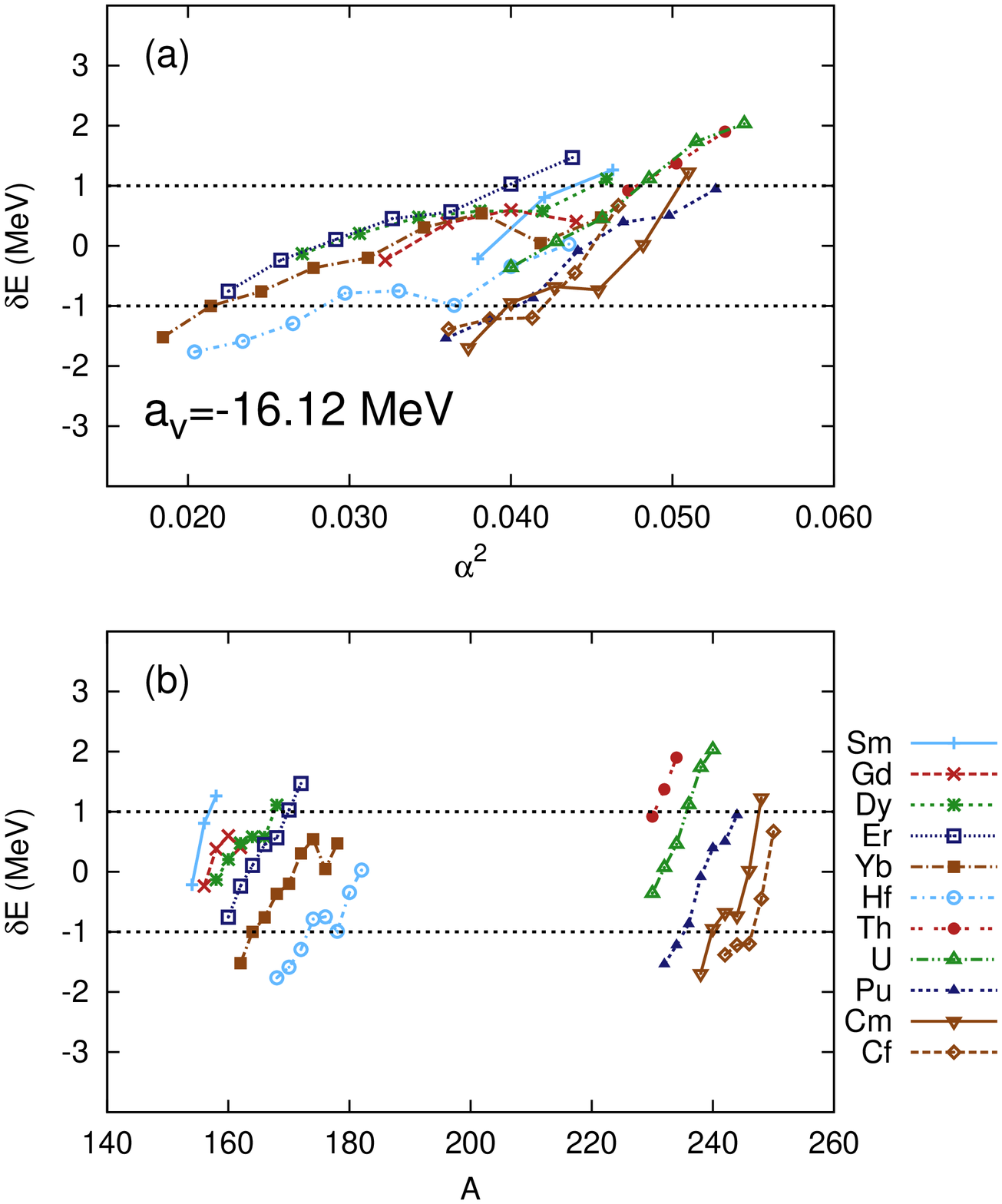}
\caption{(Color online)
Same as in Fig.~\ref{FigF}, for the point-coupling effective interaction with volume
energy $a_v=-16.12$ MeV.
}
\label{FigJ}
\end{figure}

\begin{figure}
\includegraphics[scale=0.7]{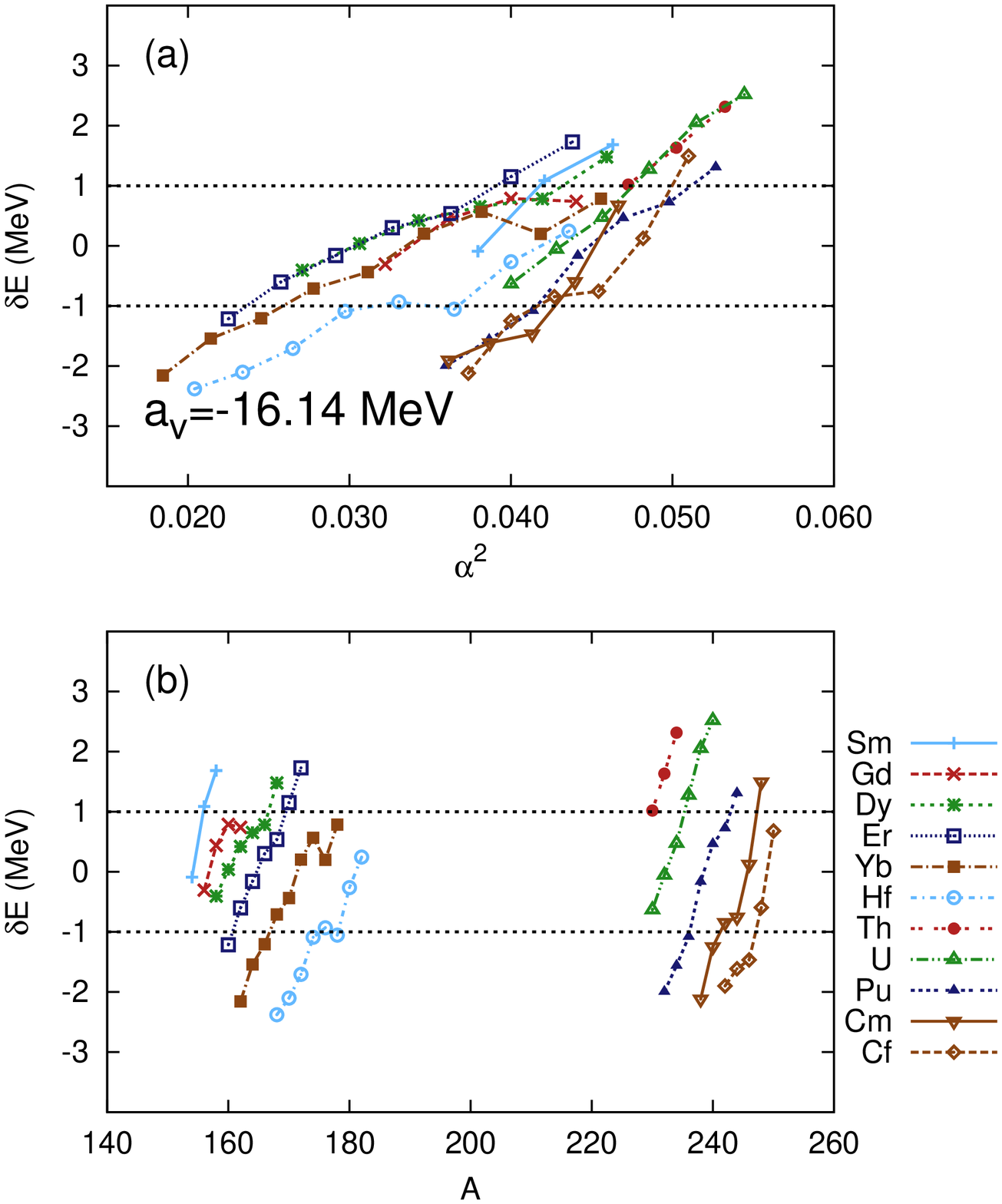}
\caption{(Color online)
Same as in Fig.~\ref{FigF}, for the point-coupling effective interaction with volume
energy $a_v=-16.14$ MeV.
}
\label{FigK}
\end{figure}

This result is illustrated in much more detail in 
Figs. \ref{FigF}-\ref{FigK}, where we display the absolute deviations of the 
calculated binding energies from the experimental values, for the 
effective interactions that correspond to each of the points
included in Fig.~\ref{FigE}.
Because these interactions have already been optimized with respect to 
$a_s$ (cf. Fig.~\ref{FigC}) and $\langle S_2 \rangle$ (cf. Fig.~\ref{FigD}), 
Figs. \ref{FigF}-\ref{FigK} show the isospin asymmetry ($\alpha^2$),  
and mass dependences of the absolute errors of calculated binding 
energies as functions of volume energy at saturation 
$a_v$. Positive deviations correspond to under-bound nuclei.
We notice that the interaction with $a_v=-16.06$ MeV (cf. Fig.~\ref{FigG}) 
not only corresponds to the lowest $\chi^2$ value, but also that it is the only one 
which does not display any visible isotopic or mass dependence of the 
deviations of calculated masses. The absolute errors for 
all 64 axially deformed nuclei in the mass regions $A\approx 150-180$ 
and $A\approx 230-250$ are smaller than $1$~MeV. With stronger 
binding in symmetric nuclear matter (i.e. by increasing the 
absolute value of $a_v$), the corresponding deviations of calculated 
binding energies become larger, and they also 
acquire a definite isotopic dependence (cf. Figs. \ref{FigH}-\ref{FigK}).
Reducing the absolute value of $a_v$ reverses the isotopic 
trend of the errors (cf. Fig.~\ref{FigF}). 
The isospin and mass dependence of binding energies shown in 
Figs. \ref{FigF} -- \ref{FigK} is 
one of the main results of the present analysis, and it illustrates
how sensitive are the calculated masses to the choice of the
nuclear matter binding energy at saturation. It also clearly shows 
why it is not possible to accurately determine the parameters of 
a nuclear energy density functional already at nuclear matter level, 
i.e. in an {\em ab initio} approach starting from empirical NN and 
NNN interactions, without additional adjustment to low-energy 
data on finite medium-heavy and heavy nuclei.

The results of Figs. \ref{FigF} -- \ref{FigK} can be compared to those
obtained with one of the most successful finite-range meson-exchange 
effective interactions: DD-ME2 (cf. Fig.~\ref{FigL}). The volume energy 
coefficient of DD-ME2  is $a_v=-16.14$ MeV, and the parameters were 
adjusted to binding energies, charge radii and neutron radii of 12 
spherical nuclei \cite{LNVR.05}. One notices both the pronounced 
isotopic and mass dependence of the deviations of binding energies 
calculated with DD-ME2. Although the span of the DD-ME2 
absolute errors is somewhat smaller than the one of the corresponding
point-coupling effective interaction with $a_v=-16.14$ MeV (cf. Fig.~\ref{FigK}), 
the meson-exchange interaction obviously underbinds most of the 64 
axially deformed nuclei, especially in the mass region $A\approx 150-180$. 
This is because DD-ME2 was adjusted to binding energies 
of spherical nuclei, and therefore implicitly includes closed-shell effects 
beyond the mean field. Virtually all self-consistent relativistic models based 
on the static mean-field approximation are 
characterized by relatively small effective nucleon masses. 
The reason is that in the RMF framework the nonrelativistic-type effective 
mass $m^*_{NR}$ \cite{DFF.05} is not independent from 
the Dirac mass $m^*_{D} = m - \Sigma_S$. 
The latter determines not only the nucleon spin-orbit potential, 
but also the binding of symmetric nuclear matter and, therefore, 
constraints the nonrelativistic-type effective mass to a rather 
narrow interval around $m^*_{NR} \approx 0.65\, m$. A small effective mass 
translates into low density of single-nucleon states around the Fermi surface. 
This is especially pronounced in magic nuclei where standard RMF models 
predict much too large energy gaps between occupied and unoccupied 
major shells. When these interactions are nevertheless forced to reproduce 
experimental binding energies of magic nuclei, i.e. when their parameters are adjusted 
to masses of nuclei like $^{132}$Sn and $^{208}$Pb, the surrounding open-shell 
nuclei are predicted under-bound, giving rise to characteristic ``arches'' of the deviations 
between theoretical and experimental binding energies \cite{BBH.06,LNVR.05}.
``Arches'' between shell closures, i.e. the over-binding of closed-shell 
nuclei relative to surrounding open-shell nuclei, characterize also most 
nonrelativistic self-consistent mean-field models, e.g. Skyrme-type 
effective interactions  \cite{BBH.06}.

It has become customary to adjust nonrelativistic and relativistic energy density 
functionals to ground-state data of magic, closed-shell nuclei. However, the 
resulting effective interactions are not often used to calculate low-lying spectra of 
spherical nuclei. These functionals are more successful 
in the description of the evolution of 
deformation, shape coexistence phenomena, rotational bands, etc. in 
deformed nuclei. This is the rationale behind the present adjustment of 
the relativistic energy density functional directly to masses of axially deformed 
medium-heavy and heavy nuclei. This procedure, of course, does not solve 
the problem of ``arches''. They are still present, but now magic, closed-shell 
nuclei are over-bound with respect to experimental binding energies.

In this section it has been shown that, among the effective 
density-dependent point-coupling interactions 
considered in the present analysis, the one with volume energy $a_v=-16.06$ MeV, surface energy $a_s=17.498$ MeV, and symmetry energy
$\langle S_2 \rangle=27.8$ MeV ($a_4 = 33$ MeV), yields best results for 
the binding energies of axially deformed nuclei in the mass regions 
$A\approx 150-180$ and $A\approx 230-250$. We will denote this interaction 
DD-PC1 (density-dependent point-coupling). In addition to the parameters 
of the isoscalar terms (set C in Table \ref{TabC}), and the strength of the 
derivative term $\delta_S = -0.815$ fm$^4$, DD-PC1 is completely specified by the two
parameters of the isovector channel: $b_{TV} = 1.836$ fm$^2$ and $d_{TV} = 0.64$. 
The total number of parameters is 10. In the next section a number of calculations 
will be performed that illustrate not only the predictive power of DD-PC1, but also 
problems in the calculation of masses of spherical nuclei. 

\begin{figure}
\includegraphics[scale=0.7]{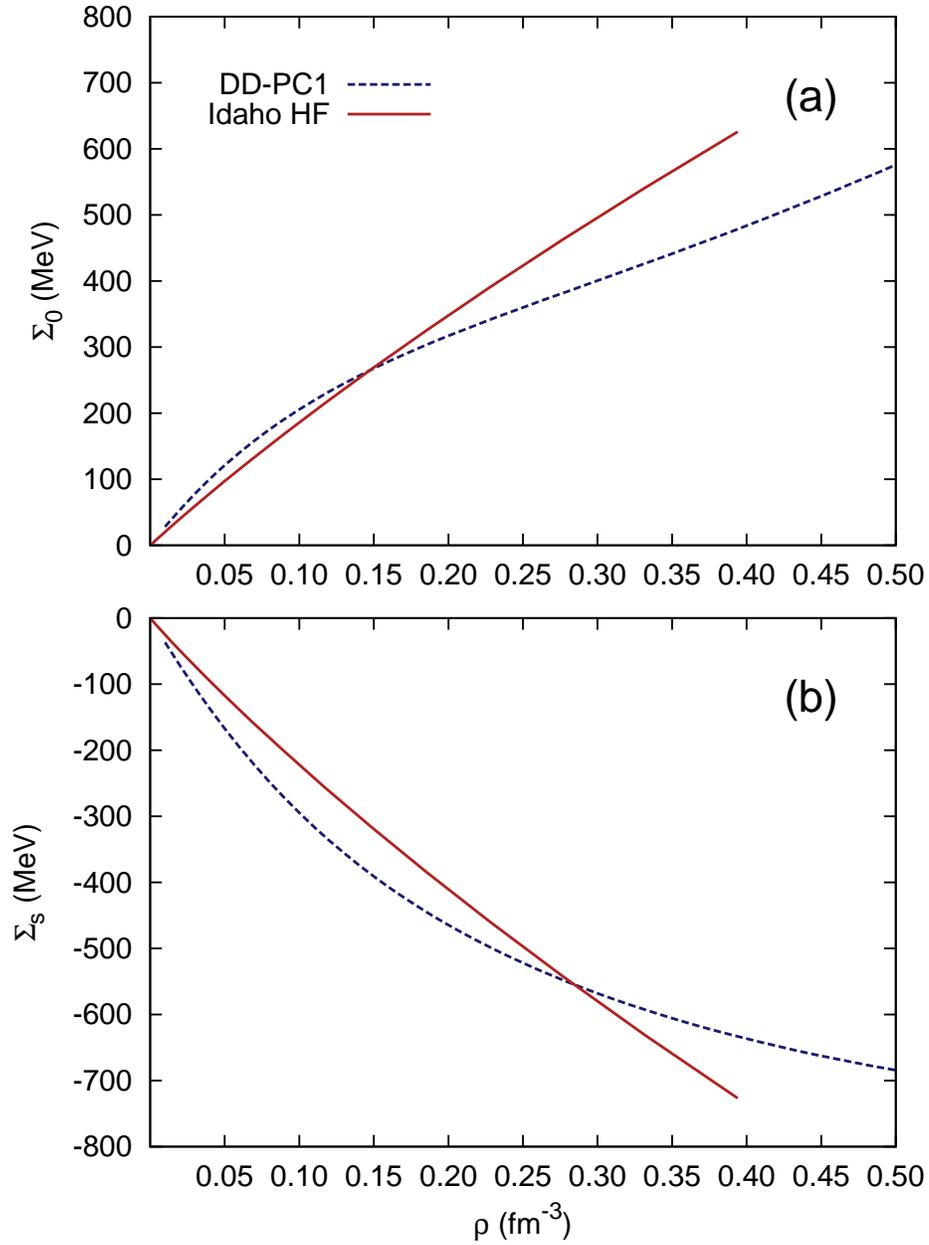}
\caption{(Color online)
Vector (upper panel) and scalar (lower panel) nucleon self-energies 
in symmetric nuclear matter as functions of the nucleon density. 
The self-energies that correspond to the phenomenological  
density functional DD-PC1 are compared with the 
Hartree-Fock self-energies \protect\cite{PF.06} calculated from 
the Idaho N$^3$LO NN-potential \protect\cite{EM.03}.
}
\label{FigZ}
\end{figure}

Finally, in Fig. \ref{FigZ} we compare the density dependence of the
DD-PC1 isoscalar vector and scalar nucleon self-energies 
in symmetric nuclear matter, with the starting approximation: 
the Hartree-Fock (HF) self-energies \cite{PF.06} calculated from 
the Idaho N$^3$LO NN-potential \cite{EM.03}. As already 
emphasized above, the analysis of Ref.~\cite{PF.06} has shown that 
at the relativistic Hartree-Fock level microscopic NN-potentials 
do not yield nuclear matter saturation. To achieve saturation of 
homogeneous symmetric matter, and to reproduce binding energies 
of finite nuclei, the self-consistent DD-PC1 mean-fields must include 
effects of short-range correlations. This necessitates an increase 
in magnitude of  the HF scalar self-energy for nucleon 
densities below $2\rho_{sat}$. 
At saturation density, in particular, this increase is 70 MeV and we 
also note the pronounced exponential dependence on density of 
the DD-PC1 self-energies, as compared to the almost linear 
density dependence of the HF Idaho self-energies. The magnitude 
of the scalar self-energy determines also the effective Dirac mass 
and, therefore, the strength of the effective nucleon spin-orbit 
potential in finite nuclei.
At low densities below $\rho_{sat}$ the HF Idaho vector self-energy 
is much less modified by the requirement of saturation and 
self-consistency. An interesting result is that, at saturation density, 
the HF Idaho and DD-PC1 vector self-energies differ by less than 3 MeV.  
At much higher nucleon densities the behavior of the DD-PC1 
self-energies has been determined by fixing the EoS to the 
point $\rho =0.56~\textnormal{fm}^{-3}$ on the 
microscopic EoS of Akmal, Pandharipande and Ravenhall \cite{APR.98}, 
and therefore it can no longer be compared with the HF Idaho self-energies.

\section{\label{secIV}Illustrative calculations}
We have performed a series of test calculations of binding energies, charge isotope 
shifts, deformations, and isoscalar and isovector giant resonances. Ground-state 
properties are calculated using the axially deformed RMF model. Pairing correlations 
are treated in the BCS approximation with constant pairing gaps determined from 
the 5-point formula:
\begin{equation}
\Delta^{(5)} (N_0) = -\frac{\Pi_{N_0}}{8} \left [E(N_0+2) - 
4E(N_0+1) + 6E(N_0) - 4E(N_0-1) + E(N_0-2) \right ] \; ,
\end{equation} 
where $E(N_0)$ denotes the experimental binding energy of a nucleus with 
$N_0$ neutrons ($Z_0$ protons), and $\Pi_{N_0} = +1~(-1)$ for 
$N_0$ even (odd).

The relativistic quasiparticle random-phase 
approximation (RQRPA) \cite{Paa.03,NVR.05} is used to calculate excitation energies 
of giant resonances in spherical nuclei. Results calculated with DD-PC1 are compared 
to available data, and with predictions of the meson-exchange interaction DD-ME2.

\begin{figure}
\includegraphics[scale=0.7]{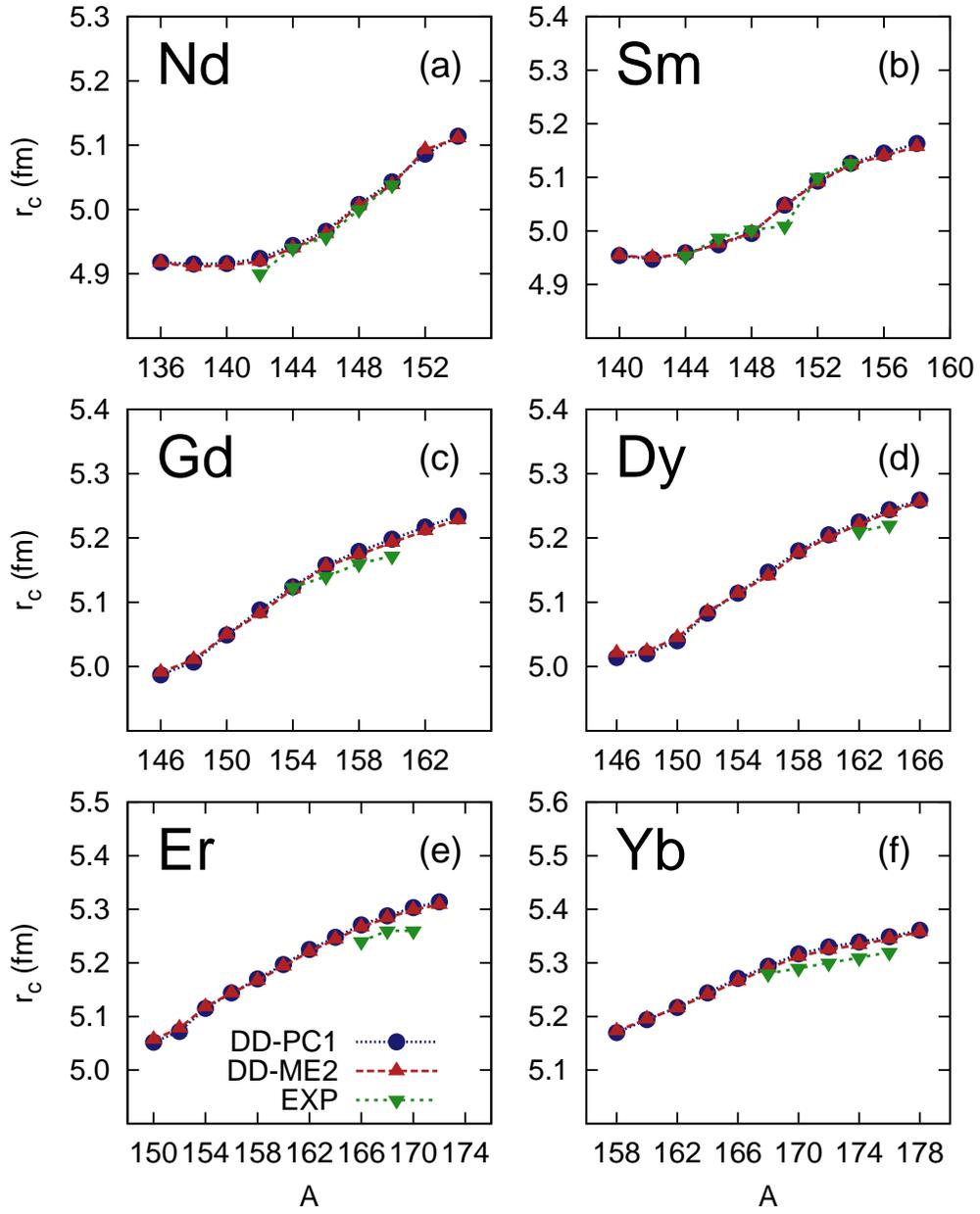}
\caption{(Color online)
Charge radii of Nd, Sm, Gd, Dy, Er and Yb isotopic chains.
The results of the RMF+BCS calculation with the DD-PC1 and 
DD-ME2 interactions are compared with
data \protect\cite{NMG.94}.
}
\label{FigM}
\end{figure}

In Fig.~\ref{FigM} the RMF+BCS predictions for charge radii
of the Nd, Sm, Gd, Dy, Er and Yb isotopic chains are 
compared with data from Ref.~\cite{NMG.94}. The charge density is
obtained by folding the theoretical point-proton density distribution with the
Gaussian proton-charge distribution. For the latter an rms radius of
0.8 fm is used, and the resulting ground-state charge radius reads
\begin{equation}
r_c = \sqrt{r_p^2+0.64}\quad {\rm fm} \;,
\end{equation}
where $r_p$ is the radius of the point-proton density distribution. 
Even though the two RMF interactions, 
meson-exchange DD-ME2 and point-coupling DD-PC1, have been 
adjusted using different procedures and to different data sets, they 
predict virtually identical charge radii for all six isotope chains. 
The theoretical values are in excellent agreement with 
data for Nd, Sm, Gd, and Dy nuclei. For the heavier 
isotopes of Er and Yb, the calculated radii are only slightly 
above the experimental values.  
Note that the parameters of DD-ME2 were tuned both to binding 
energies and charge radii of spherical nuclei, whereas only 
experimental masses of deformed nuclei have been used to 
adjust the interaction DD-PC1. Of the nuclides that belong to the 
six isotopic chains shown in Fig.~\ref{FigM}, only those with 
$N\geq 92$ are included in the data set of 64 deformed nuclei 
used to determine the density functional DD-PC1. 

\begin{figure}
\includegraphics[scale=0.7]{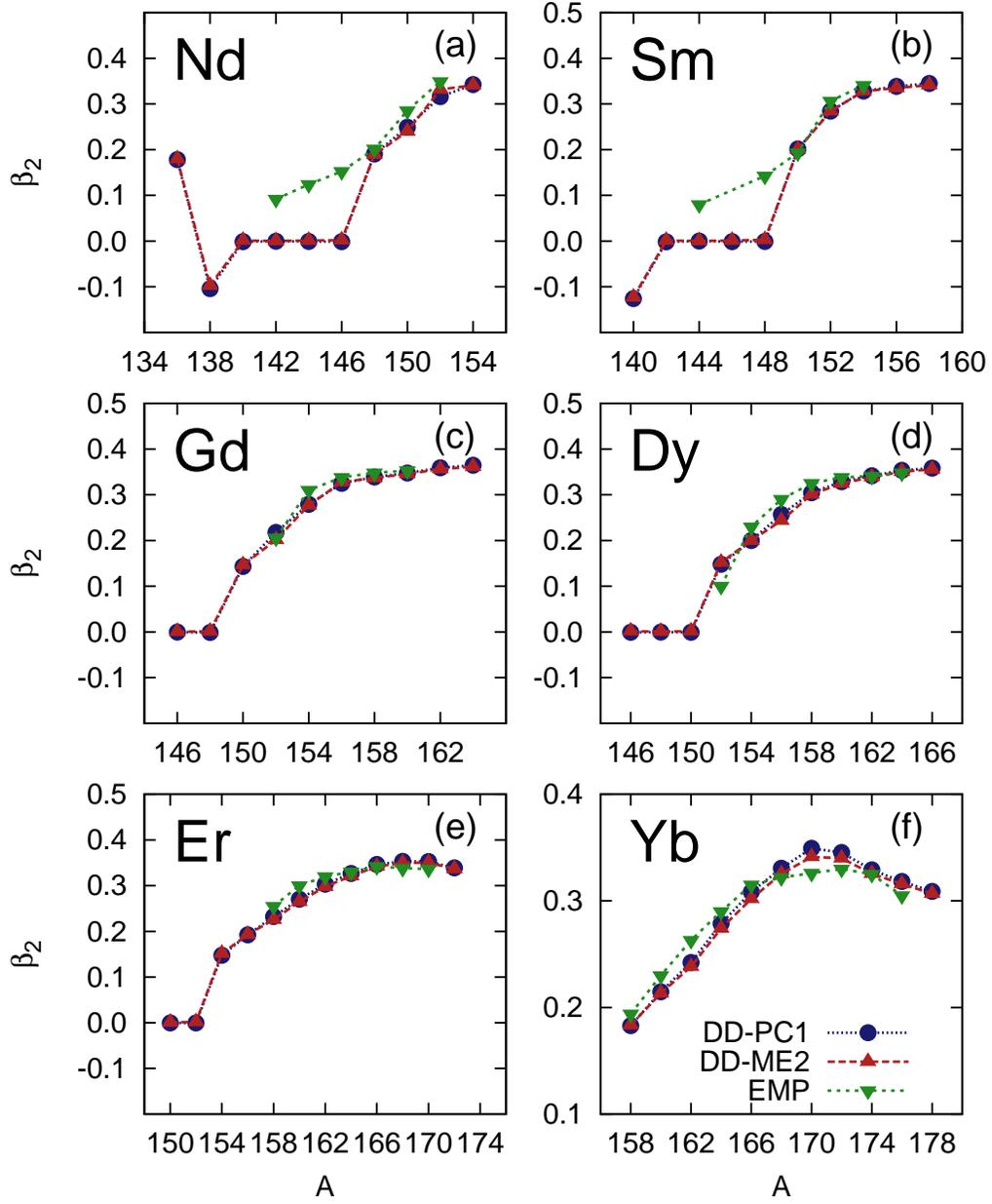}
\caption{(Color online)
DD-PC1 and DD-ME2 predictions for the ground-state 
quadrupole deformations $\beta_2$ of the Nd, Sm, 
Gd, Dy, Er and Yb isotopes, in comparison
with empirical values \protect\cite{RNT.01}.}
\label{FigN}
\end{figure}

The ground-state quadrupole deformation parameters $\beta_2$ 
are calculated according to the prescription of Ref.~\cite{LQ.82}.
The theoretical predictions for the quadrupole deformation 
parameters of Nd, Sm, Gd, Dy, Er and Yb isotopes 
are displayed in Fig.~\ref{FigN}, in comparison with 
the empirical values from Ref.~\cite{RNT.01}.
Also in this case DD-ME2 and DD-PC1 predict identical 
ground-state shapes for all six isotopic chains and  
reproduce not only the global trend of the data, but also the saturation
of quadrupole deformation for heavier isotopes. The only discrepancy 
is found around the $N=82$ closed shell in Nd and Sm isotopes, where 
both interactions predict spherical ground states, whereas data indicate 
that these nuclei are slightly prolate deformed, probably with soft 
potential energy surfaces. Shape coexistence structures (spherical-deformed, 
or prolate-oblate shapes), and soft potential surfaces in general, cannot 
quantitatively be described on a mean-field level. The description of 
coexisting shapes must include long-range correlations, and 
necessitates angular momentum projection 
and configuration mixing \cite{NVR.06a,NVR.06b}, not considered in this work.
For heavier Nd and Sm isotopes, however, the predictions of both DD-ME2 and 
DD-PC1 are again in excellent agreement with empirical prolate deformations.  

\begin{figure}
\includegraphics[scale=0.7]{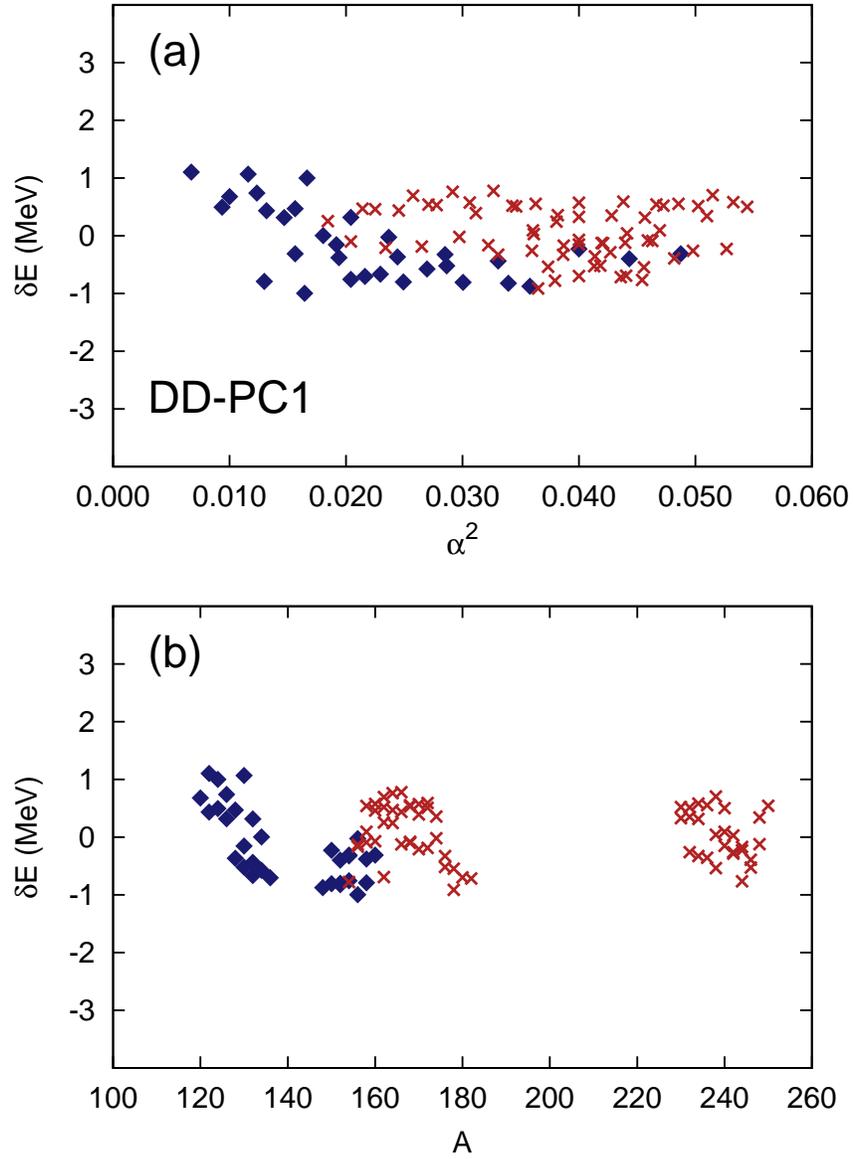}
\caption{(Color online)
Absolute deviations of the DD-PC1 binding energies 
from experimental 
values of deformed nuclei in the mass regions $A \approx 120-130$, 
$A\approx 150-180$ and $A\approx 230-250$, 
as functions of the asymmetry coefficient (upper panel), and 
mass number (lower panel).
}\label{FigO}
\end{figure}

Even though DD-PC1 is not constructed 
with the idea of being used as a mass formula, 
nevertheless this functional must also be tested in the calculation 
of binding energies. We consider the cases of deformed and 
spherical nuclei separately. As a first test Fig.~\ref{FigO} shows the
absolute deviations of the DD-PC1 binding energies 
from experimental values of deformed nuclei in the mass regions 
$A \approx 120-130$, $A\approx 150-180$ and $A\approx 230-250$, 
as functions of the asymmetry coefficient and 
mass number. Cross symbols denote the 64 nuclei used to adjust 
the parameters of DD-PC1, and correspond to the deviations already 
shown in Fig.~\ref{FigG}. Additional deformed nuclei 
that have not been used in the fit, are represented by diamond symbols. 
We include about 20 nuclei in the 
mass region $A \approx 120-130$, and 12 more around mass 
$A\approx 150-160$. In this way the predictions for binding energies 
are extrapolated to a lower mass region not included in the fit and, 
even more importantly, to lower values of the asymmetry parameter. 
The overall agreement with data is very good, and the absolute 
deviations from experiment are contained in the interval $\pm 1$ MeV.

\begin{figure}
\includegraphics[scale=0.7]{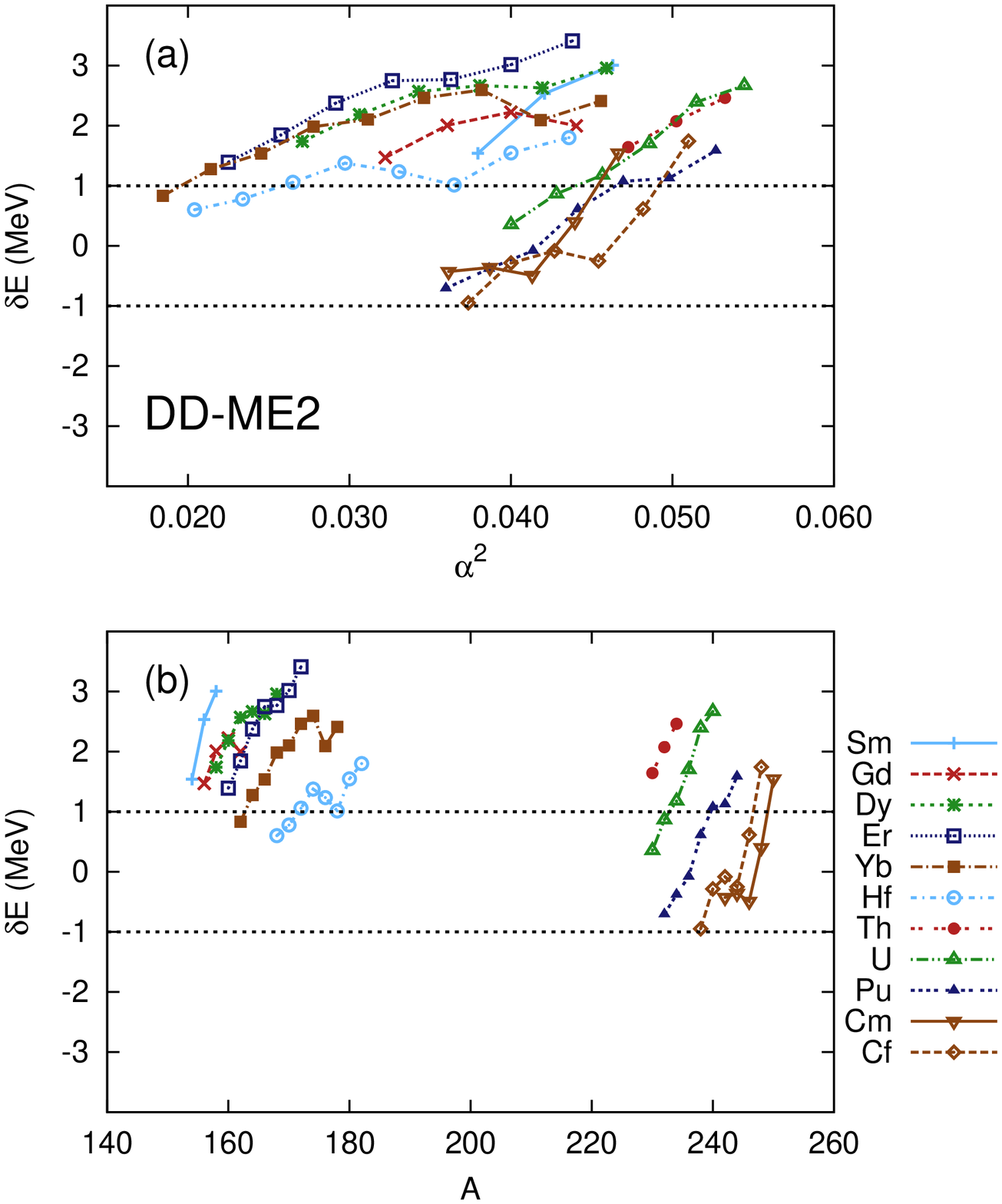}
\caption{(Color online) 
Absolute deviations of the calculated binding energies from the experimental 
values of the 64 axially deformed nuclei listed in Table \ref{TabD}, 
as functions of the asymmetry coefficient (upper panel), and 
mass number (lower panel). Lines connect nuclei that 
belong to the isotopic chains shown in the legend. 
The theoretical binding energies are calculated using the
meson-exchange effective interaction DD-ME2 \cite{LNVR.05}.
}
\label{FigL}
\end{figure}

\begin{figure}
\includegraphics[scale=0.7]{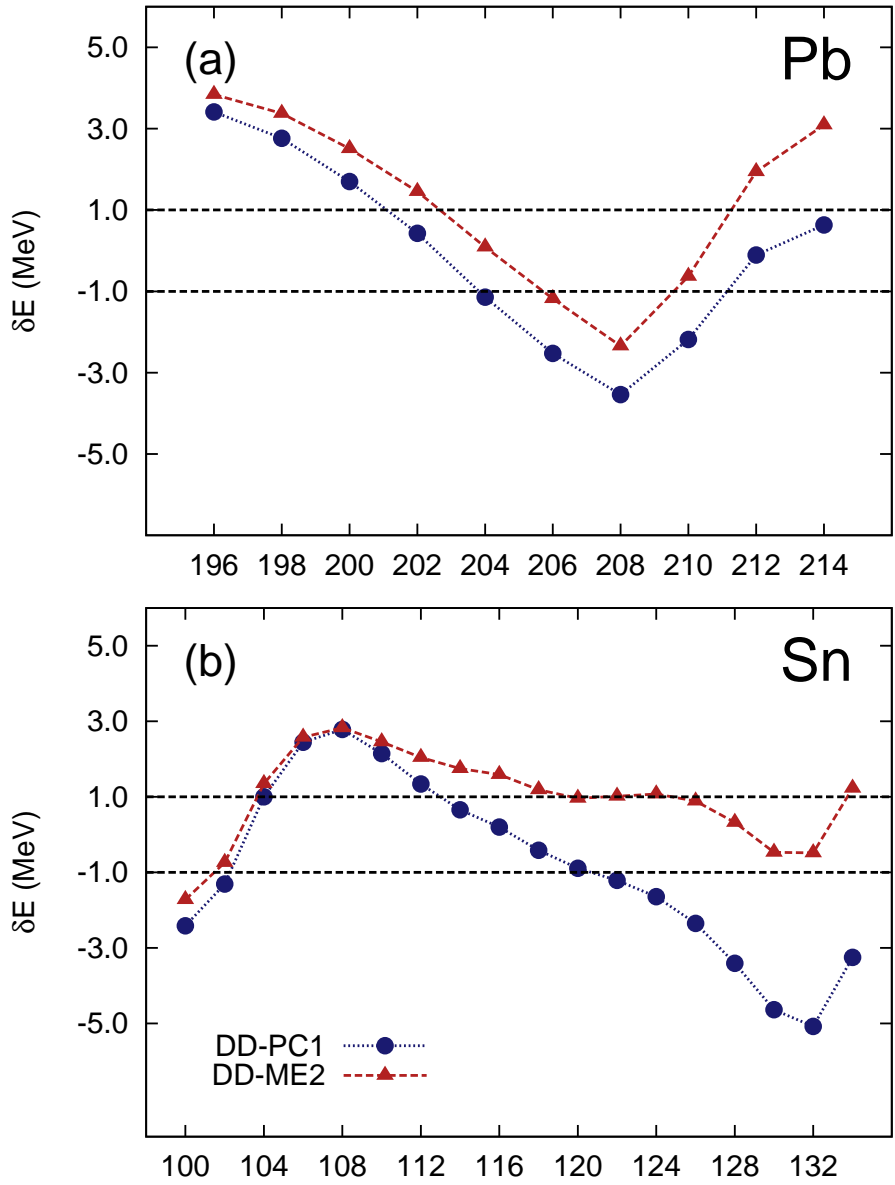}
\caption{(Color online)
Absolute deviations of the calculated binding energies from experimental 
values for the Pb (upper panel) and Sn (lower panel) isotopic chains, 
as functions of the mass number.  The theoretical binding energies are 
calculated using the RMF+BCS model with the point-coupling
effective interaction DD-PC1, and the finite-range meson exchange 
interaction DD-ME2.}
\label{FigP}
\end{figure}

In the case of spherical closed-shell nuclei, the variance between calculated 
masses and the corresponding experimental values is somewhat larger. 
This is illustrated in Fig.~\ref{FigP} where, for the Pb and Sn isotopic 
chains, we plot the absolute deviations of the calculated binding energies 
from data as functions of the mass number. The binding energies  
calculated using the RMF+BCS model with the functional DD-PC1 
are also compared to those obtained with the meson-exchange 
interaction DD-ME2. The latter, like most modern self-consistent 
mean-field nonrelativistic and relativistic interactions, was adjusted to 
reproduce the binding energies of doubly closed-shell nuclei, including  
$^{132}$Sn and  $^{208}$Pb. In addition, of the isotopes shown in 
Fig.~\ref{FigP}, the set of nuclei used to fine-tune the parameters 
of DD-ME2 included also $^{116}$Sn, $^{124}$Sn, $^{204}$Pb 
and $^{214}$Pb. The resulting binding energies of the two isotopic 
chains are, of course, in better agreement with data than those 
predicted by the density functional DD-PC1. Because it has been 
tailored to masses of deformed nuclei, DD-PC1 necessarily overbinds 
spherical closed-shell nuclei. The situation is actually not as bad as 
the comparison with the experimental mass of $^{132}$Sn might 
indicate. This isotope, with a deviation of 5.21 MeV, is in fact 
the worst case of those nuclei that we have calculated so far. For 
instance, $^{16}$O is calculated overbound by 0.72 MeV, 
$^{48}$Ca by 1.51 MeV, $^{208}$Pb by 3.51 MeV, etc. Although the 
interaction DD-ME2 predicts masses of spherical nuclei closer 
to data, it underbinds most deformed nuclei  (cf. Fig.~\ref{FigL}). 

\begin{figure}
\includegraphics[scale=0.7,angle=270]{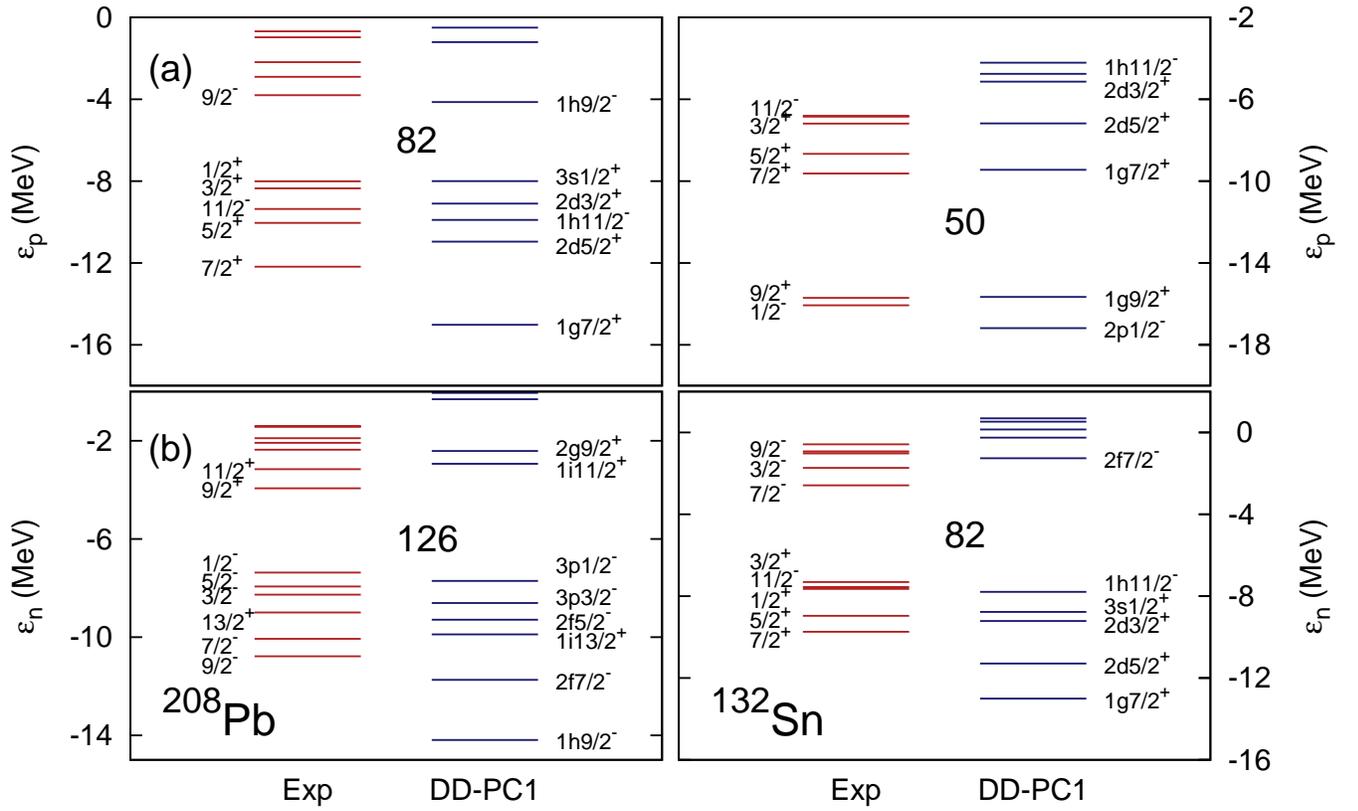}
\vspace{1cm}
\caption{(Color online) Comparison between experimental (left) and DD-PC1 (right) 
single-nucleon spectra of protons (upper panel) and neutrons (lower panel), 
for $^{208}$Pb and $^{132}$Sn. The experimental spectra are from 
Ref.~\cite{Isak.02}.
}
\label{FigR}
\end{figure}

The origin of the additional binding in closed-shell nuclei can be found in the 
predicted shell structure. For the two particular examples of 
$^{208}$Pb and $^{132}$Sn, this effect is illustrated in Fig.~\ref{FigR}, 
which shows the comparison between experimental and DD-PC1
single-nucleon spectra of protons (upper panel) and neutrons (lower panel). 
The experimental spectra correspond to the single-nucleon separation 
energies of Ref.~\cite{Isak.02}. Note that single-nucleon orbitals 
are solutions of the relativistic Kohn-Sham equations and
the corresponding eigenvalues, introduced just as Lagrange multipliers, 
do not have a directly observable physical interpretation, i.e. 
they cannot exactly be identified with nucleon separation energies \cite{Dr.Gro}.
For the last few occupied orbitals close to the Fermi surface, however, 
the Kohn-Sham eigenvalues approximately correspond to physical single-nucleon 
energies. As with most self-consistent mean-field models \cite{BHR.03}, 
the calculation reproduces the overall structure and ordering of single-nucleon 
levels, but not the level density around major shell gaps. Because of the low 
effective nucleon mass, the magnitude of the spherical shell gaps between 
occupied and unoccupied states is overestimated. As Fig.~\ref{FigR} clearly 
shows, the theoretical occupied levels are on the average considerably deeper
than the corresponding empirical single-nucleon states, and this effect gives rise 
to the overbinding that characterizes masses of spherical 
nuclei calculated with nonrelativistic and relativistic mean-field models. 
We note that a similar analysis for the nonrelativistic Skyrme interaction Sly4 
was carried out in Ref.~\cite{BBH.06}. 

This effect is particularly pronounced for the $N=82$ neutron gap in $^{132}$Sn. 
Not only is the theoretical gap much larger than the empirical one, but the 
functional DD-PC1, like many other mean-field interactions, does not reproduce 
the empirical sequence of neutron levels below $N=82$. The last occupied level  
should be 2d$_{3/2}$, whereas DD-PC1 places this orbital below 1h$_{11/2}$ 
and 3s$_{1/2}$. This increases the magnitude of the gap, and the net result 
is the additional binding of heavy Sn isotopes shown in Fig.~\ref{FigP}. If 
an interaction with a low effective nucleon mass, like DD-PC1 or DD-ME2, is 
nevertheless tuned to the masses of $^{132}$Sn and $^{208}$Pb, most 
deformed nuclei in between will be underbound, as shown for DD-ME2 in 
Fig.~\ref{FigL}.

Even though it will not be considered in the present work, we would like to 
point out that to enhance the nonrelativistic effective (Landau) 
nucleon mass, the functional 
must go beyond the static mean-field approximation, and include 
momentum-dependent (energy-dependent in stationary systems) nucleon 
self-energies. In the meson-exchange RMF framework, in particular, 
this has been achieved by including in the effective Lagrangian 
a particular form of coupling between meson fields and the 
derivatives of the nucleon fields \cite{Typ.03,Typ.05,Tom.07}. 
This leads to a linear momentum dependence of the scalar and 
vector self-energies in the Dirac equation for the in-medium nucleon. 
Although this extension of the standard mean-field framework is 
phenomenological, it is nevertheless based on Dirac-Brueckner 
calculations of in-medium nucleon self-energies, and consistent with 
the relativistic optical potential in nuclear matter, extracted from elastic 
proton-nucleus scattering data. In this way it was possible to increase 
the effective (Landau) mass to  $m^*_{NR} \approx 0.8\, m$. An additional 
enhancement of the effective nucleon mass in finite nuclei is  
caused by the coupling of single-nucleon levels to low-energy 
collective vibrational states \cite{Mah.85}, an effect which goes 
entirely beyond the mean-field approximation. In the RMF framework 
the coupling of single-nucleon states to low-energy phonons, 
and the resulting increase of the effective mass, were recently 
explored in Ref.~\cite{LR.06}.

\begin{figure}
\includegraphics[scale=0.7]{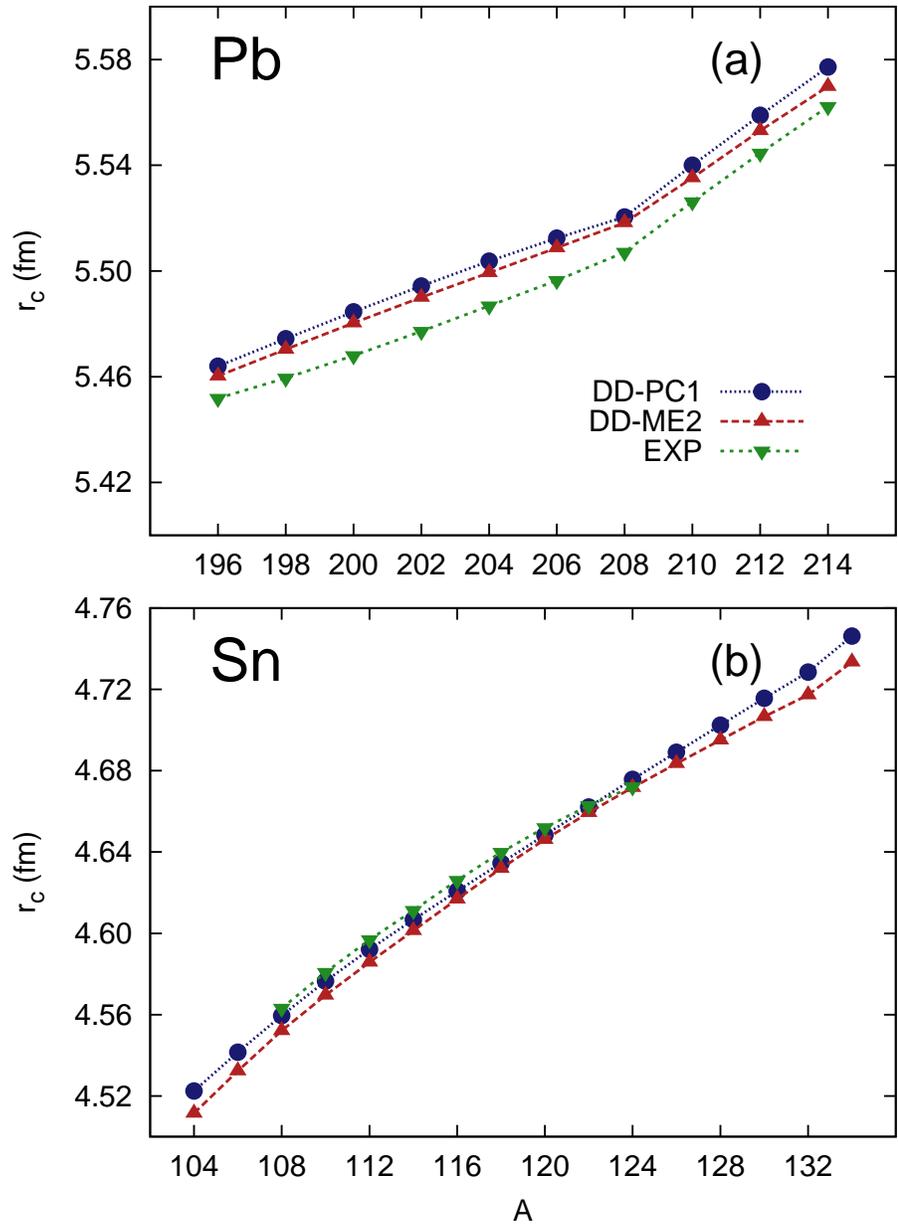}
\caption{(Color online) RMF+BCS model predictions for the charge radii of 
Pb (upper panel) and Sn (lower panel) isotopes, 
calculated with the DD-PC1 and DD-ME2 effective interactions, 
and compared with data \protect\cite{NMG.94}.
}
\label{FigS}
\end{figure}

\begin{figure}
\includegraphics[scale=0.7]{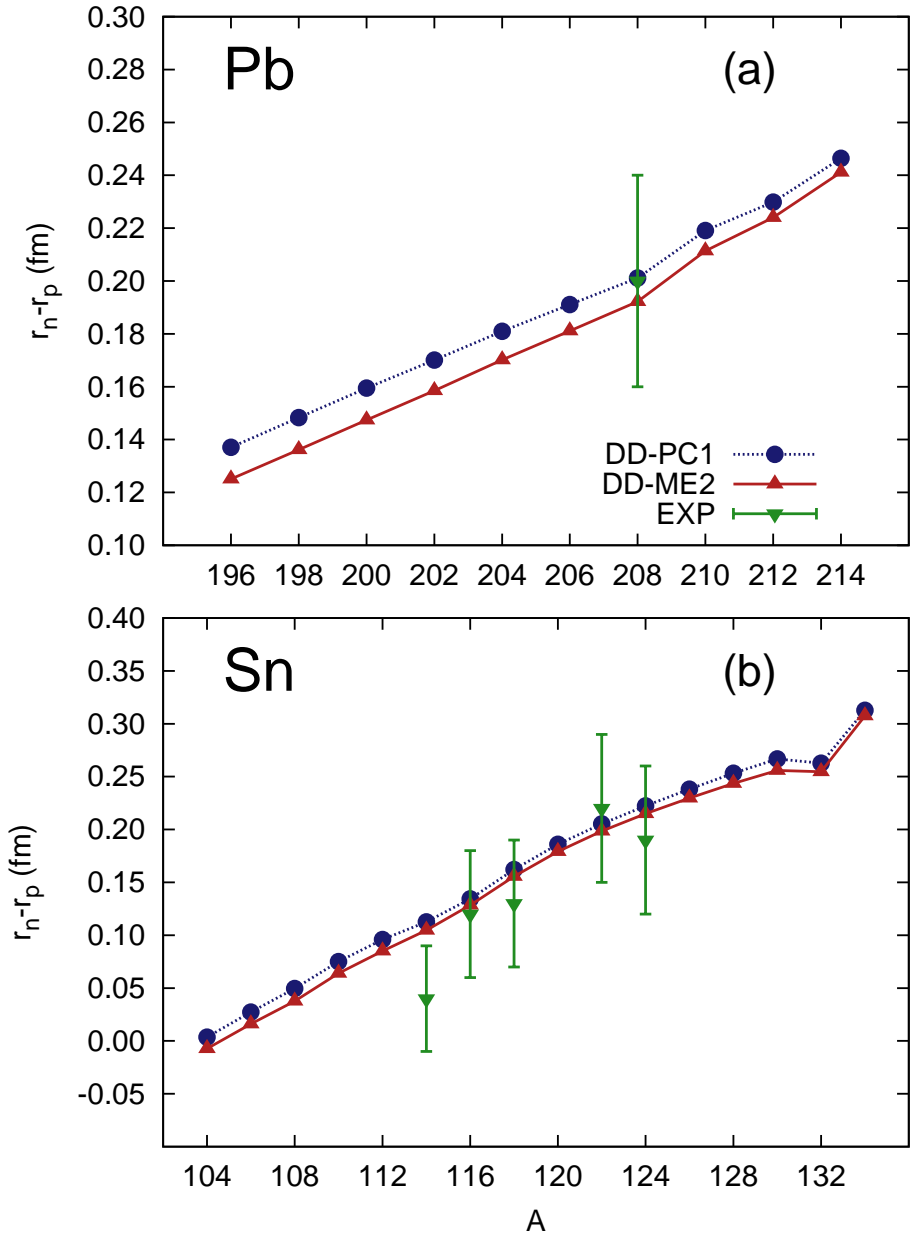}
\caption{(Color online) RMF+BCS model predictions for the 
differences between the neutron
and proton $rms$ radii of Pb (upper panel) and Sn (lower panel) isotopes, 
calculated with the DD-PC1 and DD-ME2 effective interactions,
in comparison with available data~\cite{Kra.99,SH.94,Kra.94}.
}
\label{FigT}
\end{figure}

The somewhat more pronounced deviations between the theoretical 
DD-PC1 and experimental masses of the Sn and Pb isotopic chains, 
do not affect the accuracy of the calculated radii of these nuclei. In 
Fig.~\ref{FigS} the RMF+BCS model predictions for the
charge radii of Pb and Sn isotopes, calculated with the 
effective interactions DD-PC1 and DD-ME2, 
are shown in comparison with empirical values \cite{NMG.94}. 
The two interactions predict virtually identical values for the charge 
radii, in excellent agreement with data for the Sn nuclei, and only 
slightly above the empirical charge radii of the Pb isotopes. We 
note again that the charge radii of  $^{116}$Sn, $^{124}$Sn, 
$^{132}$Sn, $^{204}$Pb,  $^{208}$Pb, and $^{214}$Pb were 
used to tune the parameters of DD-ME2, whereas DD-PC1 
has only been adjusted to masses. A similar result is also 
obtained for the thickness of the neutron skin in Pb and Sn nuclei. 
In Fig.~\ref{FigT} the DD-ME2 and DD-PC1 results  for the 
differences between the neutron and proton $rms$ radii are
compared with available data~\cite{Kra.99,SH.94,Kra.94}. 
Although the experimental uncertainties are large, 
both interactions nicely reproduce the isotopic trend of neutron 
radii in Sn nuclei, and predict values in very good agreement with 
the empirical values of neutron skin thickness. These data, 
however, were specifically used to adjust the isovector channel 
of the DD-ME2 interaction, whereas in the case of DD-PC1 this 
level of agreement is achieved with the choice of the asymmetry 
energy at saturation $a_4 = 33$ MeV.

\begin{figure}
\includegraphics[scale=0.7,angle=270]{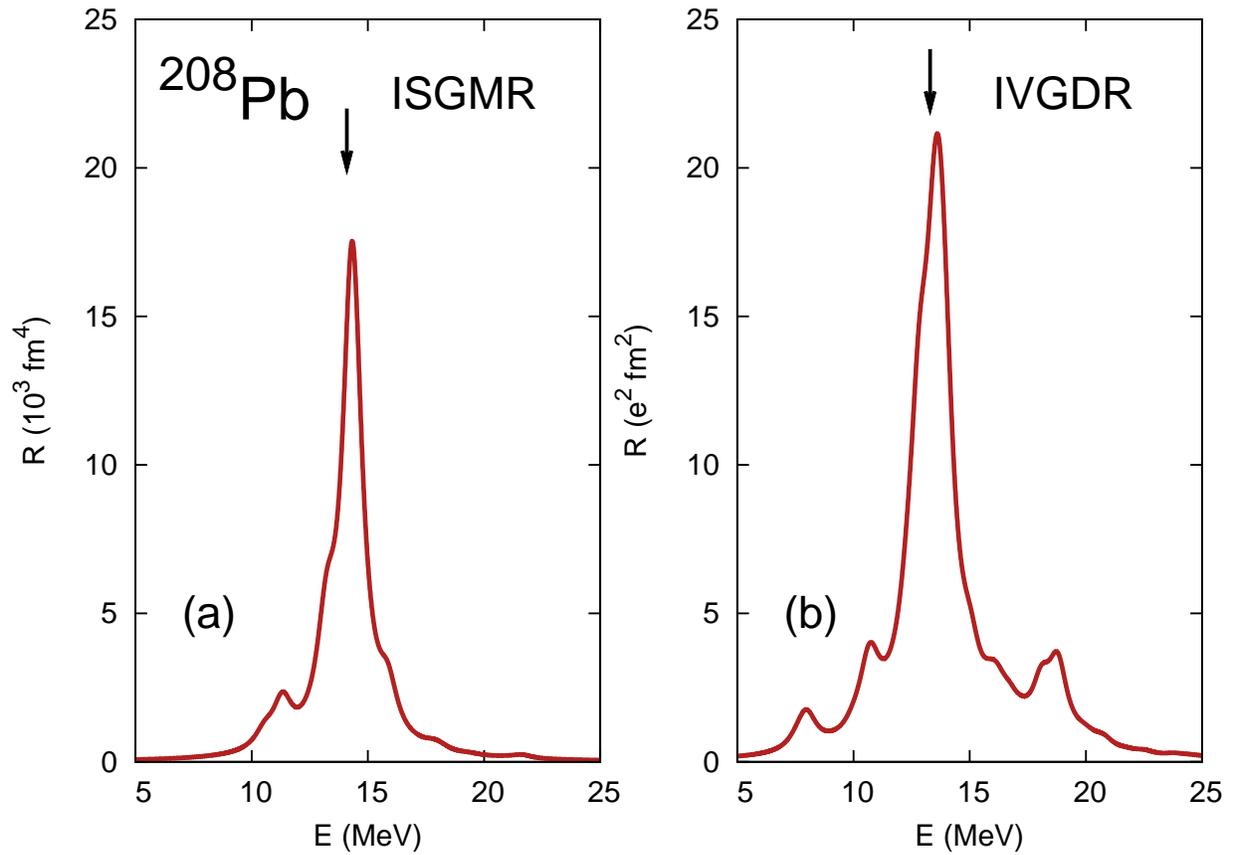}
\vspace{2cm}
\caption{(Color online) The isoscalar monopole (left panel) and isovector dipole 
(right panel) strength distribution in $^{208}$Pb
calculated with the relativistic RPA using the 
effective interaction DD-PC1. The experimental
excitation energy are denoted by arrows: 
$13.96\pm 0.2$~\cite{Young.04} for the giant monopole resonance, 
and $13.3\pm 0.1$~\cite{Rit.93} for the giant dipole resonance, 
respectively.
}
\label{FigW}
\end{figure}

Another very important test of self-consistent mean-field models are excitation 
energies of collective modes and, in particular, giant multipole resonances.
Using the relativistic (Q)RPA \cite{Paa.03,NVR.05} with the DD-PC1 functional, 
we have therefore carried out few representative calculations of giant 
resonances in spherical nuclei. The RQRPA is fully self-consistent, i.e. 
both in the particle-hole and particle-particle channels the same interactions 
are used in the equations that determine the ground state of a nucleus, and as 
residual interactions in the matrix equations of RQRPA. The RQRPA 
configuration space includes also the Dirac sea of negative energy states.
In adjusting the parameters of DD-PC1 we took into account the 
results of Refs.~\cite{NVLR.08}, where it has been shown that,
to reproduce the experimental excitation energies of 
isoscalar giant monopole resonances (ISGMR) in spherical nuclei,
relativistic point-coupling interactions require a
nuclear matter compression modulus of $K_{nm}\approx 230$ MeV, 
somewhat lower than the values typically used for meson-exchange relativistic
interactions \cite{VNR.03}, and within the range of values used by 
modern nonrelativistic Skyrme interactions. In Ref.~\cite{VNR.03} 
it was also shown that data on the isovector giant 
dipole resonance (IVGDR) constrain the 
range of the nuclear matter symmetry energy at
saturation to 31 MeV $\leq a_4 \leq$ 35 MeV, and in 
this work we have used $a_4 = 33$ MeV for all point-coupling 
interactions. For $^{208}$Pb the RRPA results for the 
isoscalar monopole and isovector dipole 
response are shown in Fig.~\ref{FigW}. 
For the multipole operator $\hat Q_{\lambda \mu}$ the 
response function $R(E)$ is defined 
\begin{equation}
R(E) = \sum_i~B(\lambda_i \rightarrow 0)~
{{\Gamma/ {2 \pi}}\over {(E - E_i)^2 + \Gamma^2/4}},
\label{Lorentz}
\end{equation}
where $| 0 \rangle$ denotes the ground state of an even-even nucleus,
$|\lambda_i \rangle$ is the corresponding $i$-th excited state of 
multipolarity $\lambda$,  
 $\Gamma$ is the width of the Lorentzian distribution, and
\begin{equation}
B(\lambda \rightarrow 0)  =  \frac{1}{2\lambda + 1} \, |
\langle 0 || \hat{Q}_\lambda || \lambda \rangle |^2.
\end{equation}
The $k$-th moment of the strength function is defined by:
\begin{equation}\label{eq:moments}
m_k(\hat{Q_\lambda}) = 
\sum_i ~ E_i^k~ |\langle \lambda_i   | \hat{Q}_\lambda | 0\rangle|^2 
\end{equation}
The ratio $m_1/m_0$ is the quantity often compared with the 
experimental excitation energy of the corresponding resonance 
although, of course, this is strictly correct only if there are no
pronounced multiple peaks within the energy interval over which the 
sumation in Eq.~(\ref{eq:moments}) is performed. In the examples considered here 
the continuous strength distributions are obtained 
by folding the discrete spectrum of R(Q)RPA states with the 
Lorentzian (cf. Eq.~(\ref{Lorentz})) of constant
width $\Gamma = 1$ MeV. 

The calculated excitation energies of the giant resonances in $^{208}$Pb
can be compared with very accurate data. For the ISGMR the calculated 
$m_1/m_0 = 14.17 $ MeV is rather 
close to the experimental value $m_1/m_0 = 13.96\pm 0.2$ MeV \cite{Young.04}. 
The relativistic RPA  peak energy of the IVGDR at 13.6 MeV is also in 
very good agreement with the experimental excitation energy  
$E^* = 13.3\pm 0.1$ MeV \cite{Rit.93}. A similar level of agreement, 
both for ISGMR and IVGDR, is also obtained with the DD-ME2 interaction, 
however we note that these data were taken into account in adjusting the
 parameters of DD-ME2 \cite{LNVR.05}.

\begin{figure}
\includegraphics[scale=0.65,angle=270]{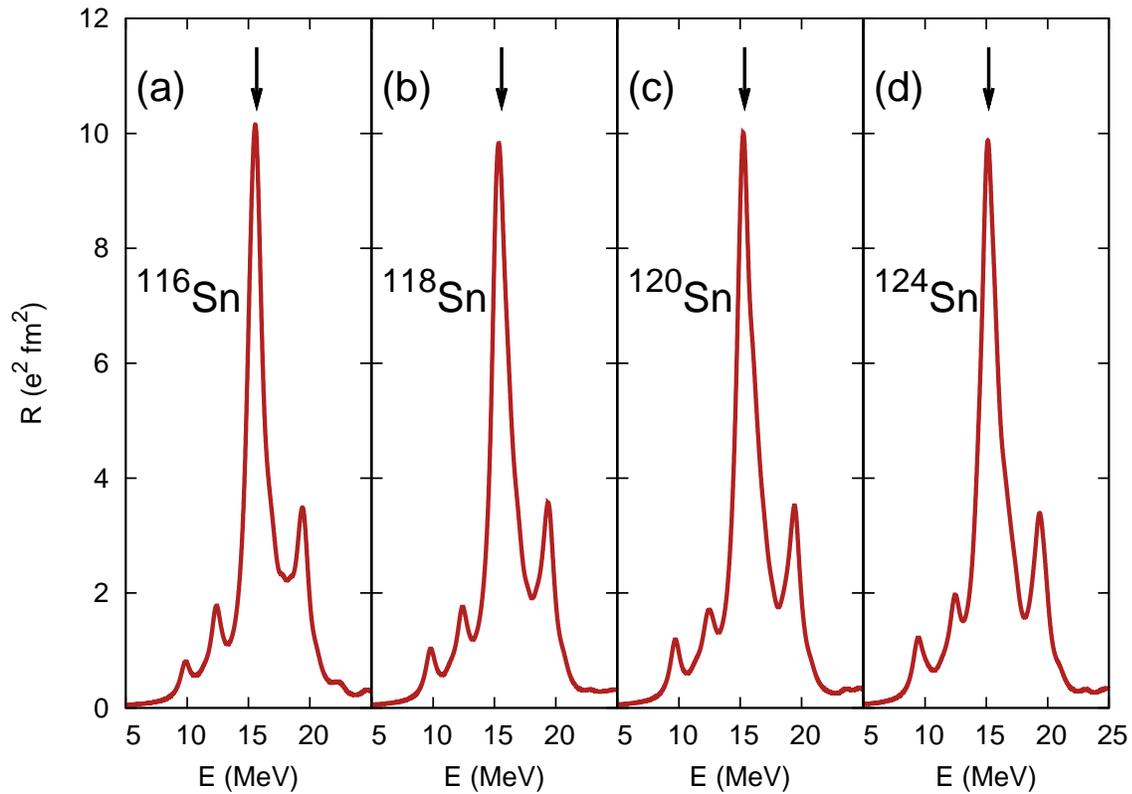}
\vspace{1cm}
\caption{(Color online) The RQRPA isovector dipole strength functions in 
$^{116,118,120,124}$Sn, calculated with the DDPC1 effective interaction.
The experimental IVGDR excitation energies~\cite{Ber.75}
are denoted by arrows.
}
\label{FigX}
\end{figure}

In Fig.~\ref{FigX} the RQRPA results for the Sn isotopes 
are compared with
data on IVGDR excitation energies~\cite{Ber.75}.
In contrast to the case of $^{208}$Pb,
the strength distributions in the region of giant resonances 
exhibit considerable fragmentation. Note, however, that the
RQRPA calculation with the DD-PC1 interaction reproduces
the isotopic dependence of the IVGDR and the
experimental excitation energies. The theoretical peak energies 
15.56 MeV ($^{116}$Sn), 15.35 MeV ($^{118}$Sn), 
15.26 MeV ($^{120}$Sn), 15.13 MeV ($^{124}$Sn), are 
in excellent agreement with the experimental values:
15.68 MeV ($^{116}$Sn), 15.59 MeV ($^{118}$Sn), 
15.36 MeV ($^{120}$Sn), 15.19 MeV ($^{124}$Sn), respectively.

\begin{figure}
\includegraphics[scale=0.65,angle=270]{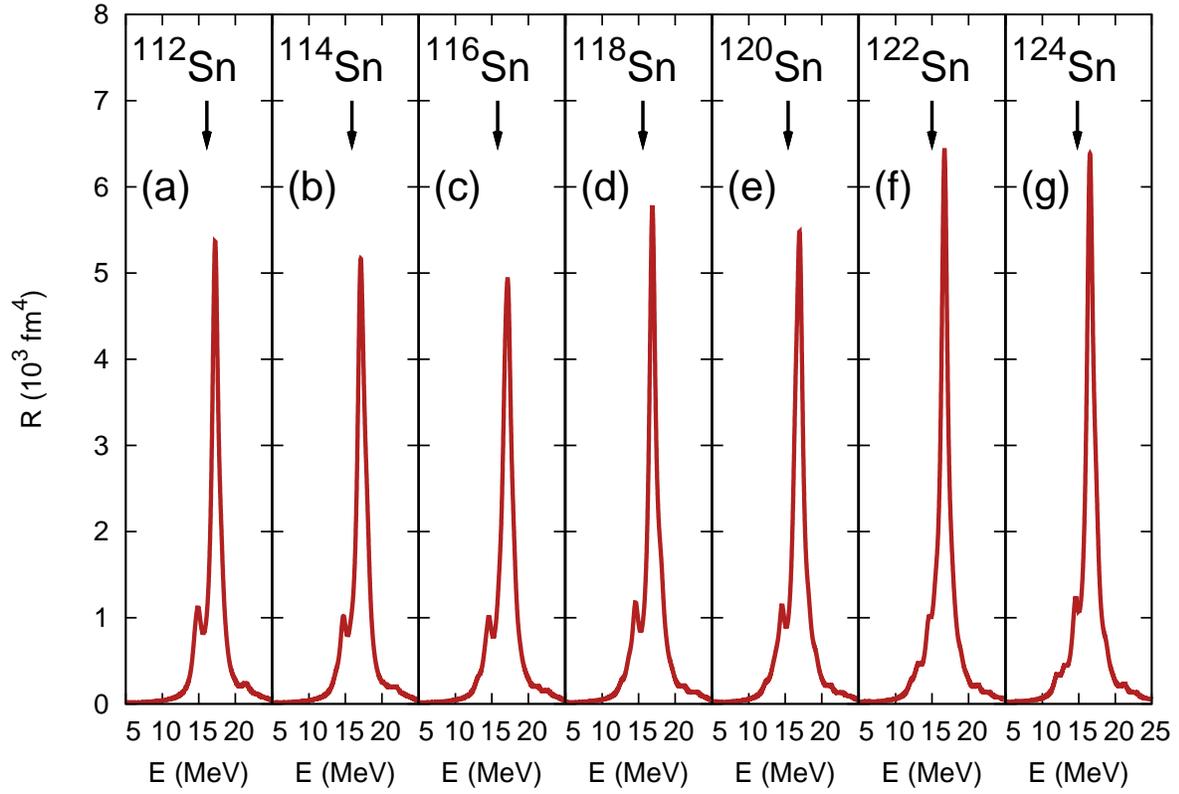}
\vspace{1cm}
\caption{(Color online) The RQRPA isoscalar monopole strength distributions in 
even-even $^{112-124}$Sn nuclei, calculated with the DDPC1 effective interaction.
Arrows denote the positions of experimental ISGMR 
excitation energies~\protect\cite{Li.07}.
}
\label{FigY}
\end{figure}

Finally, in Fig.~\ref{FigY} we display the RQRPA isoscalar monopole strength 
functions for the chain of even-even Sn isotopes: $^{112-124}$Sn. 
The evolution of ISGMR in Sn isotopes can be compared with very recent 
data from Ref.~\cite{Li.07}. In general, the theoretical excitation 
energies $E_0=m_1/m_0$, when compared with the corresponding 
experimental values evaluated in the energy interval between 
10.5 MeV and 20.5 MeV, are systematically somewhat higher:
between 0.8 MeV and 1 MeV. This result might indicate that 
the value of the nuclear matter compression modulus for 
this functional could actually be chosen $K_{nm} < 230$ MeV. 
However, Sn isotopes are much lighter than $^{208}$Pb, and 
one can expect that the calculated ISGMR will be more affected 
by the surface incompressibility, a quantity that we have not 
attempted to determine in this work.


\section{\label{secV}Summary and concluding remarks}		     

The principal goal of modern nuclear structure theory is to 
build a consistent microscopic framework that will describe 
ground-state properties, nuclear excitations and reactions
at a level of accuracy comparable with experimental results, 
and provide reliable predictions for systems very far from stability, 
including data for astrophysical applications that are not accessible 
in experiments. At present the only viable approach to a comprehensive 
description of arbitrarily heavy nuclear systems, including vast regions 
of short-lived nuclei with extreme isospin values and extended 
nucleonic matter, is the one based on nuclear energy density functionals.

In this work we have explored a particular class of relativistic nuclear energy 
density functionals in which only nucleon degrees of freedom are explicitly 
used in the construction of effective interaction terms. Short-distance correlations, 
as well as intermediate and long-range dynamics, are effectively taken into account 
in the nucleon density dependence of the strength functionals of second-order contact 
interactions in an effective Lagrangian. Starting from microscopic nucleon 
self-energies in nuclear matter, a phenomenological ansatz for the density-dependent 
coupling functionals has been formulated and the corresponding parameters adjusted
in self-consistent mean-field calculations of masses of 64 axially deformed nuclei in 
the mass regions $A\approx 150-180$ and $A\approx 230-250$. 

The relationship between global properties of the nuclear matter equation of state 
(EoS) and the corresponding predictions for nuclear masses has been analyzed 
in detail. Ground states of deformed nuclei have been calculated 
in the self-consistent mean field approximation by employing 
sets of effective interactions with different values of the volume $a_v$, 
surface $a_s$, and symmetry energy $a_4$ in nuclear matter, whereas 
empirical constraints have been used for the nuclear matter saturation density, compression modulus, and Dirac effective mass. The calculated masses are 
not particularly sensitive to the saturation density, and previous relativistic 
RPA calculations of excitation energies of isoscalar giant monopole and 
isovector giant dipole resonances in finite nuclei, as well as results for the 
neutron skin thickness have been used to determine the nuclear matter 
compression modulus and symmetry energy, respectively. 

One of the important results of the careful analysis of deviations between 
calculated and experimental masses (mass residuals), is the pronounced 
isospin and mass dependence of the residuals on the nuclear matter volume 
energy at saturation. To reduce the absolute mass residuals to less than 
1 MeV, and to contain their mass and isotopic dependence, strict constraints 
on the value of $a_v$ must be met. The narrow window of allowed values 
of the volume energy cannot be determined microscopically already at the 
nuclear matter level, but rather results from a fine-tunning of the parameters 
of the energy density functional to experimental masses. 
Calculated binding energies and charge radii are also sensitive to the 
choice of the surface coefficient $a_s$ that determines the 
surface energy and surface thickness of semi-infinite nuclear matter. 
In the functional considered in the present work, these quantities are 
controlled by the strength of the derivative isoscalar-scalar coupling interaction. 
The range of allowed values of the strength parameter determined by data is 
in very good agreement with estimates obtained from microscopic calculations 
of inhomogeneous nuclear matter. 

The optimal energy density functional (DD-PC1) determined in a multistep 
parameter fit to the masses of 64 axially deformed nuclei, has been further tested 
in calculations of properties of spherical and deformed medium-heavy and heavy 
nuclei, including binding energies, charge radii, deformation parameters, neutron skin 
thickness, and excitation energies of giant multipole resonances. Results have been 
compared with available data, and with predictions of one of the most successful 
finite-range meson-exchange relativistic effective interactions: DD-ME2. In general, 
a very good agreement with data has been obtained except, perhaps, for the effect 
of overbinding of spherical closed-shell nuclei. DD-PC1, like virtually all relativistic 
mean-field models, is characterized by a relatively low effective nucleon mass and, 
when adjusted to masses of deformed nuclei, it overbinds spherical closed-shell 
systems. The well known problem of ``arches'' of mass residuals between shell 
closures could be addressed by a functional that goes beyond the static  
mean-field approximation, but this approach has not been considered in the 
present model. Very good results have been obtained for the excitation energies 
of giant monopole and dipole resonances in spherical nuclei, calculated with 
the relativistic quasiparticle random-phase approximation based on the DD-PC1 
functional. The agreement with data confirms the choice of the nuclear matter 
compressibility and symmetry energy for DD-PC1.

The total number of parameters in the functional DD-PC1 is 10, like in most 
non-relativistic Skyrme-type density functionals. Note, however, that the 
effective Lagrangian of DD-PC1 contains only four interaction terms 
except, of course, the Coulomb term (cf. Eq.~(\ref{Lagrangian})). The 10 
parameters determine the density dependence of the strength functionals and 
reflect the complex nuclear many-body dynamics. We also emphasize that, 
because the high-density behavior of the corresponding nuclear matter EoS 
has been adjusted to a microscopic equation of state extensively used in 
studies of high-density nucleon matter and neutron stars, DD-PC1 should also 
be tested in astrophysical applications.

This work is part of a program to develop an universal relativistic energy density 
functional to be used in studies of of the evolution of shell structure, deformation, 
shape coexistence phenomena, rotational bands, etc. in transitional and 
deformed nuclei. In the first step the parameters of the density functional have been 
adjusted to reproduce binding energies of a large set of axially deformed nuclei. 
In the continuation of this program we plan to build a functional, based on DD-PC1, 
to be used in the new relativistic model that employs the generator coordinate method (GCM) to perform configuration mixing of angular-momentum and
particle-number projected mean-field wave functions. However, if 
rotational energy corrections and quadrupole fluctuations are treated explicitly 
in the GCM framework, they should not at the same time implicitly be included in 
the functional, i.e. through parameters adjusted to data that already include 
correlations. Therefore, starting from DD-PC1, 
the parameters of this new functional can be adjusted to pseudodata, 
obtained by subtracting correlation effects from experimental masses and radii. 
The resulting energy density functional will be tested in relativistic GCM model 
studies of shell evolution, deformations, shape coexistence and shape phase transitions. 

\bigskip\bigskip
\leftline{\bf ACKNOWLEDGMENTS} We thank O. Plohl and 
C. Fuchs for providing the Hartree-Fock nucleon self-energies  
based on the Idaho N$^3$LO NN-potential.
This work was supported in part by MZOS - project 1191005-1010, 
and by the DFG cluster of excellence \textquotedblleft Origin and
Structure of the Universe\textquotedblright\ (www.universe-cluster.de).
\bigskip


\clearpage
\clearpage
\clearpage
\clearpage
\clearpage
\clearpage
\clearpage
\clearpage
\clearpage
\clearpage
\clearpage
\clearpage
\clearpage
\clearpage
\clearpage
\clearpage
\clearpage
\clearpage
\clearpage
\clearpage
\clearpage
\clearpage
\clearpage
\clearpage
\begin {table}[tbp]
\begin {center}
\caption{Isoscalar parameters Eq.~(\ref{ansatz}) of the
point-coupling effective interactions with 
volume energy coefficients: 
$a_v=-16.02$ MeV (set A), $a_v=-16.04$ MeV (set B), $a_v=-16.06$ MeV (set C),
$a_v=-16.08$ MeV (set D) , $a_v=-16.10$ MeV (set E), $a_v=-16.12$ MeV (set F),
$a_v=-16.14$ MeV (set G) and $a_v=-16.16$ MeV (set H).
}
\bigskip
\begin {tabular}{ccccccccc}
\hline
\hline                          
 parameter & set A & set B  &  set C & set D & set E & set F & set G & set H
 \\ \hline
$a_S$ (fm$^2$)  & $-10.0220$  & $-10.0332$  & $-10.0462$ & $-10.0855$                     & $-10.0951$ &   $-10.1051$ & $-10.1137$ & $ -10.1220 $\\
$b_S$ (fm$^2$)  & $-9.1781$ & $-9.1666$  & $-9.1504$ & $-9.0623$                              & $-9.0539$ &   $-9.0436$ & $-9.0384 $ & $-9.0307 $\\
$c_S$ (fm$^2$)  & $-6.2799$ & $-6.3541$  & $-6.4273$  & $-6.4878$                                                     & $-6.5611$ &   $-6.6336$ & $-6.7065 $ & $-6.7786 $\\
$d_S$       & $1.3585$ & $1.3654$ & $1.3724$   & $1.3806$  
& $1.3872$ & $1.3938$  & $1.4001 $ & $1.4065$\\ \hline 
$a_V$  (fm$^2$)& $5.9020$ & $5.9108$   & $5.9195$   & $5.9262$   
& $5.9348$ &   $5.9431$ & $5.9513$ & $5.9594$\\
$b_V$  (fm$^2$)& $8.8711$ & $8.8687$   & $8.8637$   & $8.8156$   
& $8.8150$ &  $8.8134$ & $8.8148$ & $8.8147$\\
$c_V$ (fm$^2$)& $0.0$          &$0.0$          & $0.0$         & $0.0$         
& $0.0$       &$0.0 $  & $0.0$ &  $0.0$ \\
$d_V$       & $0.6548$ & $0.65676$   & $0.6584$   & $0.6547$   
& $0.6568$ & $0.6587$ & $0.6610 $ & $0.6630$ \\ 
\hline
\hline	
\end{tabular}
\label{TabC}
\end{center}
\end{table}
\begin {table}[tbp]
\begin {center}
\caption{The binding energies of the isotopic chains 
$62\leq Z\leq 72$ and $90\leq Z\leq 98$ have been 
used to adjust the parameters of relativistic
point-coupling effective interactions. $N_{min}$ and
$N_{max}$ denote the corresponding ranges
of neutron number in even-even nuclides.}
\bigskip
\begin {tabular}{cccccccccccc}
\hline
\hline         
$Z$ & $62$ & $64$ & $66$ & $68$ & $70$ & $72$ & $90$ & $92$ & $94$ & $96$ & $98$ \\
\hline
$N_{min}$ & $92$ & $92$ & $92$ & $92$ & $92$ & $72$ & $140$ & $138$ & $138$ & $142$ & $144$ \\
$N_{max}$& $96$ & $98$ & $102$ & $104$ & $108$ & $110$ & $144$ & $148$ & $150$&$152$&$152$\\
\hline \hline                 
\end{tabular}
\label{TabD}
\end{center}
\end{table}

\begin{thebibliography}{999}

\bibitem {BHR.03}M. Bender, P.-H. Heenen, and P.-G. Reinhard, Rev. Mod. Phys.
{75}, 121 (2003).

\bibitem {VALR.05}D. Vretenar, A. V. Afanasjev, G. A. Lalazissis, and P. Ring,
Phys. Rep. {409}, 101 (2005).

\bibitem{KS.65} W. Kohn and L. J. Sham,
	Phys. Rev.  {140},  A1133 (1965).
	
\bibitem{Kohn.99} W. Kohn, Rev. Mod. Phys. {71}, 1253 (1999).  

\bibitem{LNP.641} 
G. A. Lalazissis, P. Ring, and D. Vretenar (Eds.), 
	\textit{Extended Density Functionals in Nuclear Structure Physics},
	Lecture Notes in Physics {641}, (Springer, Heidelberg 2004)  
	
\bibitem{SW.86} B. D. Serot and J. D. Walecka, Adv. Nucl. Phys. 16, 1 (1986).

\bibitem{SW.97} B. D. Serot and J. D. Walecka, Int. J. Mod. Phys. E 6, 515 (1997).

\bibitem{FS.00} R. J. Furnstahl and B. D. Serot, 
	Comments Nucl. Part. Phys. 2, A23 (2000).
	
\bibitem{Joe.05} J. N. Ginocchio, Phys. Rep. 414, 165 (2005).
 
\bibitem {FKV.04}P. Finelli, N. Kaiser, D. Vretenar, and W. Weise, 
	Nucl. Phys. A 735, 449 (2004).

\bibitem {FKV.06}P. Finelli, N. Kaiser, D. Vretenar, and W. Weise, 
	Nucl. Phys. A 770, 1 (2006).

\bibitem{FKW.05} S. Fritsch, N. Kaiser, and W. Weise, Nucl. Phys. A750, 259 (2005). 

\bibitem{NVR.06a} T. Nik\v{s}i\'{c}, D. Vretenar, and P. Ring, 
	Phys. Rev. C 73, 034308 (2006).

\bibitem{NVR.06b} T. Nik\v{s}i\'{c}, D. Vretenar, and P. Ring, 
	Phys. Rev. C 74, 064309 (2006).

\bibitem{BBH.06} M. Bender, G. F. Bertsch, and P.-H. Heenen, 
	Phys. Rev. C 73, 034322 (2006).
	
\bibitem{BBH.04} M. Bender, G. F. Bertsch, and P.-H. Heenen, 
	Phys. Rev. C 69, 034340 (2004).

\bibitem{BBH.06a} M. Bender, P. Bonche, and P.-H. Heenen, 
	Phys. Rev. C 74, 024312 (2006).

\bibitem{Cle.07} E. Cl{\' e}ment et al., 
		Phys. Rev. C 75, 054313 (2007).
			
\bibitem{NVLR.07} T. Nik\v{s}i\'{c}, D. Vretenar, G. A. Lalazissis, and P. Ring,
		Phys. Rev. Lett. 99, 092502 (2007).

\bibitem {MNH.92}D. G. Madland, B. A. Nikolaus, and T. Hoch, 
	Phys. Rev. C 46, 1757 (1992).

\bibitem {Hoch.94}T. Hoch, D. Madland, P. Manakos, T. Mannel, B.A. Nikolaus,
	and D. Strottman, Phys. Rep. 242, 253 (1994).

\bibitem {FML.96}J. L. Friar, D. G. Madland, and B. W. Lynn, 
	Phys. Rev. C 53, 3085 (1996).

\bibitem {RF.97}J. J. Rusnak and R. J. Furnstahl, Nucl. Phys. A 627, 495 (1997).

\bibitem{BMM.02}
	T. B{\"u}rvenich, D. G. Madland, J. A. Maruhn, and P.-G. Reinhard, 
	Phys. Rev. C 65,  044308  (2002).

\bibitem{HKL.01} F. Hofmann, C. M. Keil, and H. Lenske, 
	Phys. Rev. C 64, 034314 (2001). 

\bibitem{TW.99} S. Typel and H. H. Wolter, Nucl. Phys. A 656, 331 (1999).

\bibitem{NVFR.02} T. Nik\v{s}i\'{c}, D. Vretenar, P. Finelli, and P. Ring, 
	Phys. Rev. C 66, 024306 (2002).

\bibitem{FLW.95} C. Fuchs, H. Lenske, and H. H. Wolter, Phys. Rev. C 52, 3043 (1995).

\bibitem{JL.98} F. de Jong and H. Lenske, Phys. Rev. C 57, 3099 (1998).

\bibitem{LNVR.05} G. A. Lalazissis, T. Nik\v{s}i\'{c}, D. Vretenar and P. Ring, 
               Phys. Rev. C 71, 024312 (2005).	
               
\bibitem{LMGZ.04} W. Long, J. Meng, N. Van Giai and S.-G. Zhou, 
	Phys. Rev. C 69 034319 (2004).
                
\bibitem{Long.06} W. Long, N. Van Giai, and J. Meng, Phys. Lett. B 640, 150 (2006).


\bibitem{PF.06} O. Plohl and C. Fuchs, Phys. Rev. C 74, 034325 (2006).

\bibitem{EM.03} D. R. Entem and R. Machleidt, Phys. Rev. C 68, 041001(R) (2003).

\bibitem{NVLR.08} T. Nik\v{s}i\'{c}, D.Vretenar, G. A. Lalazissis, and P. Ring, 
	Phys. Rev. C 77, 034302 (2008).
	
\bibitem{APR.98} A. Akmal. V. R. Pandharipande, and D. G. Ravenhall,
	Phys. Rev. C 58, 1804 (1998).
	
\bibitem{BSU.05} G. F. Bertsch, B. Sabbey, and M. Uusn\"akki, 
	Phys. Rev. C 71, 054311 (2005).
	
\bibitem{Fur.02} R. J. Furnstahl, Nucl. Phys. A 706, 85 (2002).	

\bibitem{VNR.03} D. Vretenar, T. Nik\v{s}i\'{c}, and P. Ring, 
	Phys. Rev. C  68,  024310 (2003).        
	
\bibitem{HF.89} D. Hofer and W. Stocker, Nucl. Phys. A 492, 637 (1989).

\bibitem{CPC} P. Ring, Y. K. Gambhir, and G. A. Lalazissis, 
	Comput. Phys. Commun. 105, 77 (1997).

\bibitem{DFF.05} E. N. E. van Dalen, C. Fuchs, and Amand Faessler, 
	Phys. Rev. Lett. 95, 022302 (2005). 
	
\bibitem{AW.03} G. Audi, A. H. Wapstra, and C. Thibault,
	Nucl. Phys. A 729, 337 (2003).

\bibitem{Paa.03} N. Paar, P. Ring, T. Nik\v si\' c and D. Vretenar,
	Phys. Rev. C 67, 034312 (2003).

\bibitem{NVR.05} T. Nik\v{s}i\'{c}, D. Vretenar, and P. Ring, 
	Phys. Rev. C 72, 014312 (2005).

\bibitem{NMG.94} E. G. Nadjakov, K. P. Marinova, and Yu. P. Gangrsky, 
      At. Data Nucl. Data Tables  56, 133 (1994).

\bibitem{LQ.82} J. Libert and P. Quentin, Phys. Rev. C {\bf 25}, 571 (1982).

\bibitem{RNT.01} S. Raman, C. Nestor, and P. Tikkanen, 
	At. Data Nucl. Data Tables 78, 1 (2001).

\bibitem{Isak.02} V. I. Isakov, K. I. Erokhina, H. Mach, 
		M. Sanchez-Vega, and B. Fogelberg,
		Eur. Phys. J. A 14, 29 (2002).

\bibitem{Dr.Gro} R. M. Dreizler and E. K. U. Gross, 
	Density Functional theory, Spinger-Verlag, 1990.

\bibitem{Typ.03} S. Typel, T. v. Chossy,  and H.H. Wolter, 
	Phys. Rev. C 67, 034002 (2003). 
	
\bibitem{Typ.05} S. Typel, Phys. Rev. C 71, 064301 (2005).

\bibitem{Tom.07} T. Marketin, D. Vretenar, and P. Ring,
		Phys. Rev. C 75, 024304 (2007).

\bibitem{Mah.85} C. Mahaux, P. F. Bortignon, R. A. Broglia, and C. H. Dasso, 
	Phys. Rep. 120, 1 (1985).		

\bibitem{LR.06} E. Litvinova and P. Ring, Phys. Rev. C 73, 044328 (2006).
	
\bibitem{Kra.99} A. Krasznahorkay et al., 
	Phys. Rev. Lett. 82, 3216 (1999).

\bibitem{SH.94} V. E. Starodubsky and N. M. Hintz,
	Phys. Rev. C 49, 2118 (1994).

\bibitem{Kra.94} A. Krasznahorkay et al.,
	Nucl. Phys. A 567, 521 (1994).
		
\bibitem{Young.04} D. H. Youngblood, Y.-W. Lui, H. L. Clark, B. 
	John, Y. Tokimoto, and X. Chen. Phys. Rev. C 69, 034315 (2004).   
	
\bibitem{Rit.93} J. Ritman et al., Phys. Rev. Lett. 70, 533 (1993). 
	      
\bibitem{Ber.75} B. L. Berman and S. C. Fultz, Rev. Mod. Phys. 47, 713 (1975).

\bibitem{Li.07} T. Li et al.,
		Phys. Rev. Lett. 99, 162503 (2007).
		
\end{thebibliography}
\end{document}